# Modeling Frequency Shifts in Small-Scale Beams with Multiple Eccentric Masses*

Hossein Darban[a, 1], Raimondo Luciano[b], Michał Basista[a]

[a] *Institute of Fundamental Technological Research, Polish Academy of Sciences, Pawińskiego 5B, 02-106 Warsaw, Poland*
[b] *Department of Engineering, University of Naples Parthenope, 80133, Naples, Italy*

## Abstract

Studying the dynamics of small-scale beams with attached particles is crucial for sensing applications in various fields, such as bioscience, material science, energy storage devices, and environmental monitoring. Here, a stress-driven nonlocal model is presented for the free transverse vibration of small-scale beams carrying multiple masses taking into account the eccentricity of the masses relative to the beam axis. The results show excellent agreement with the experimental and numerical data in the literature. New insights into the frequency shifts and mode shapes of the first four vibrational modes of stress-driven nonlocal beams with up to three attached particles are presented. The study investigates the inverse problem of detecting the location and mass of an attached particle based on natural frequency shifts. The knowledge acquired from the present study provides valuable guidance for the design and analysis of ultrasensitive mechanical mass sensors.

## 1. Introduction

### 1.1. Micro- and Nanostructures

Micro- and Nanoelectromechanical Systems (MEMS and NEMS) represent one of the most significant and effective technological advancements of recent decades. Nowadays, these miniaturized systems are integrated into numerous devices in our daily lives. They are cost-effective and easy to manufacture while offering high precision and accuracy in multitasking. Due

**[1]- *Corresponding author:*** *Hossein Darban*; ***E-mail address:*** hdarban@ippt.pan.pl; ***Tel.:*** *(+48) 22 826 12 81*

to their excellent characteristics, MEMS and NEMS have been extensively utilized in various fields of science, technology, and industry, including transportation, communication, manufacturing, environmental monitoring, healthcare, energy, and aerospace [1]. Currently, they are widely used in the fabrication of microfluidic devices, capacitive micromachined ultrasonic transducers, sensors, switches, and inkjet printer heads, among others. The use of MEMS and NEMS continues to expand, introducing new possibilities to our world every day.

A recent example of a miniaturized system with crucial functionality is the micro-shutter arrays used in NASA's James Webb Space Telescope. Another example is the mechanical mass sensors based on micro- and nanobeams. These compact sensors provide highly sensitive and real-time mass detection capabilities [2]. Information about the mass can contribute to advancing concepts related to the structure and nature of biological entities. Additionally, mass sensors play a crucial role in medical diagnosis [3], the monitoring of air pollution [4], the evaluation of hydrogen storage capacity in energy storage devices [5], and the study of fundamental phenomena in surface science, such as phase transitions and diffusion [6]. These sensors are typically employed in either static mode, measuring deflection caused by surface stresses resulting from the binding of entities, or dynamic mode, measuring frequency shifts following the attachment of ultrasmall masses and molecules.

### 1.2. Modeling Techniques

The presence of size dependence distinguishes the study of micro- and nanostructures from their large-scale counterparts. The size dependence has been repeatedly observed in experimental works such as those reported in [7,8] and is due to several physical reasons. One factor contributing to this size dependence is the intrinsic length of the material, such as the lattice parameter and grain size, becoming comparable to the geometrical dimensions in micro- and nanostructures. Other physical causes for the emergence of size-dependent structural behavior include the rearrangement of atoms or molecules near the surface and long-range influences from dislocations and voids. While atomistic models [9] such as molecular dynamics and molecular mechanics can accurately predict the mechanical response of micro- and nanostructures, they are inherently burdened by complex formulations and substantial computational costs. Therefore, the scientific community highly appreciates studying miniaturized structures without excessive mathematical complexity using the continuum mechanics-based approaches. This modeling approach necessitates appropriate constitutive laws to account for micro- and nanoscale size effects in the formulation, something the classical continuum mechanics models do not account for.

Numerous pioneering nonclassical continuum mechanics-based models have been developed to explore the size-dependent mechanical behavior of micro- and nanostructures. The frequently utilized theories are based on the modified couple stress, the strain gradient, and the nonlocal theories. For instance, both linear and nonlinear Bernoulli–Euler and Timoshenko beam models, based on the modified couple stress theory, are developed in the seminal papers [10–13] to study the statics and dynamics of micro- and nanobeams. Structural theories based on the strain gradient elasticity theory are formulated in [14–18] to study different size-dependent mechanical problems in micro- and nanoscale tubes, beams, and plates. Review papers [19,20] thoroughly examine theoretical analyses and numerical formulations based on several nonclassical continuum mechanics approaches to study the linear and nonlinear behavior of miniaturized beams, plates, and shells, with a focus on bending, buckling, and dynamic characteristics.

Among several pioneering nonclassical continuum mechanics-based approaches, the nonlocal elasticity theory stands out as one of the most efficient for studying the size-dependent response of structures with small-scale dimensions, e.g., [21–24]. In this approach, the constitutive relation is assumed to have a nonlocal nature, meaning that instead of using a pointwise stress-strain relation, the stress or strain of a given point is influenced by the stress or strain of all points in the domain. A weighting kernel function is required to control the intensity of nonlocality in a given problem. This idea was introduced by Eringen [25] and has since been used by many others to study the size-dependent response of structures with micro- and nanoscale dimensions, e.g., [26–29]. The Eringen nonlocal theory is also known as the strain-driven nonlocal theory since the stress at a given point is defined as an integral convolution of the strains at all points of the body and a kernel function. As far as the differential form of the Eringen nonlocal theory is concerned, the formulation may become mathematically ill-posed for some problems with technological importance, such as a cantilever beam with an end force [30].

More recently, an innovative type of nonlocal theory, namely the stress-driven nonlocal theory, which is free of inconsistencies, has gained much attention [31]. The idea is to define the strain at any reference point in terms of the stresses at all points of the body. Since models based on the stress-driven approach always yield a well-posed formulation, the approach has been widely applied to study various problems of micro- and nanostructures, e.g. [32,33]. The capability of the stress-driven nonlocal theory to model dynamic and static experiments at small scales is illustrated in [34]. Important contributions are made [35–40], where stress-driven nonlocal theories are formulated for studying problems with non-smooth fields. In such problems, the solution can be obtained by decomposing the domain at the locations where the discontinuities exist. The discontinuities can be in the form of concentrated loads, kinematic constraints, cracks, or attached masses at the interior parts of the structure.

## 1.3. Micro- and Nanostructures with Attached Masses

Dynamics analysis of small-scale structures with attached masses is important for mass sensing applications. This problem has been explored through experimental, numerical, and theoretical approaches. Some of the experimental works on the micro- and nanomechanical mass sensors for applications related to the environmental, chemical, and biological fields are discussed in the review article [2]. In [41], a novel class of nanoplate-based mass sensors with corner point supports is studied using a nonlocal elasticity theory. A novel sandwich mass sensor composed of (i) a smart core made of a functionally graded magneto-electro-elastic nanofilm and (ii) graphene faces, is introduced and modeled using a nonlocal strain gradient theory and a first-order shear deformation plate theory [42]. In [43], the frequency shifts of a single-layered graphene sheet, caused by the attachment of particles in the presence of a magnetic field, are examined using nonlocal Kirchhoff-Love plate theory. The analysis of nanobeams and nanorods with single-point masses is conducted through the application of the modified strain gradient theory [44,45]. The study in [46] investigates the large-amplitude, size-dependent dynamics of a functionally graded microcantilever with an intermediate spring support and a tip mass, employing the modified couple stress theory. Exploration of the coupled axial-flexural vibration in cantilever mass nanosensors is carried out in [47], employing a two-phase local/nonlocal elasticity approach. The axial vibration behavior of mass sensors based on single-walled carbon nanotubes is studied in [48] using Eringen's nonlocal elasticity theory. It is demonstrated that the axial vibration behavior of single-walled carbon nanotubes can be effectively utilized in mass sensors with zeptogram-scale mass sensitivity. These are just a few examples of theoretical studies in the literature on the size-dependent structural dynamics of mass sensors based on nonlocal theories.

The application of the stress-driven theory of nonlocal elasticity for a comprehensive exploration of size effects in micro- and nanomechanical mass sensors is very rare. One contribution is outlined in [49], focusing solely on the scenario of a nanocantilever with a single attached particle positioned at the free end. However, in reality, multiple particles may simultaneously bind to different positions along the sensor. Addressing this gap, [38] employed the stress-driven theory of nonlocal elasticity to investigate the impact of size dependence on the precision of micro- or nanomechanical mass sensors. To maintain generality, the study considered the scenario of a mass sensor with multiple attached particles at arbitrary locations. Nevertheless, the model presented in [38] is developed under the assumption that the attached particles are relatively small compared to the sensor dimensions, thus neglecting the eccentricity of the particles with respect to the axis line of the sensor.

In reality, the particles are attached to the surfaces of the sensors. Therefore, the center of mass of an attached particle is always located at a distance from the axis of the sensor. In many cases, where the size of a biological entity is negligible compared to the size of a relatively large mechanical mass sensor with a length of a few hundred microns, the effect of the eccentricity of the attached particle may be neglected. However, micro- and nanomechanical mass sensors show great promise, especially when they have smaller dimensions, lengths below 10 micrometers, and thicknesses on the order of tens of nanometers. This is because only at these diminutive ranges do the sensors exhibit enhanced sensitivity to detect even single cells and viruses [3]. For instance, gold-coated silicon nanomechanical cantilever mass resonators with femtogram-level mass sensitivity were fabricated in [50], measuring 2 to 6 microns long and 50 to 100 nanometers thick. Considering that the size of numerous biological entities, such as viruses, bacteria, and proteins, becomes comparable to the dimensions of ultrasensitive micro- and nanomechanical mass sensors, neglecting the eccentricity of attached particles in analyzing such sensors becomes questionable. For instance, cantilever mass sensors with lengths as small as 15 microns and thicknesses of 320 nanometers were used in [51] to detect *Escherichia coli* cells with lengths, widths, and thicknesses of approximately 1.43, 0.73, and 0.35 microns, respectively. Similarly, silicon cantilevers with a length of 6 microns and a thickness of 150 nanometers were used [52] to detect the mass of a nonpathogenic insect baculovirus with a length of 0.5 microns and a thickness of 25 nanometers. Similar studies are reported in [53,54], where *Listeria* bacteria, proteins, and vaccinia virus particles are detected using ultrasensitive mechanical silicon mass sensors with dimensions comparable to those of the attached particles. Therefore, it is important to account for the effect of the eccentricity of attached masses when analyzing micro- and nanomechanical mass sensors.

Furthermore, incorporating the eccentricity of attached particles into sensor modeling not only captures the underlying physics more accurately but also enhances our comprehension and enriches the existing knowledge of ultrasensitive mass sensors.

### 1.4. Objective

The objective of this work is to formulate a model for the size-dependent free transverse vibration of micro- and nanobeams with multiple eccentric masses using the stress-driven nonlocal theory. The model formulated in this paper extends the work in [38] in three main aspects: (i) considering the eccentricity of the attached masses, which enhances the accuracy of the results for sensing large entities, (ii) presenting results for clamped-clamped beams, commonly used in practical applications as mass sensors, and (iii) exploring the inverse problem of detecting the intensity of the mass and its location based on the frequency shifts. The axial vibration caused by the eccentricity of the attached masses is not considered, assuming that this mode of vibration does not affect the transverse vibration of the beam. This assumption, previously used in the literature (e.g., in [55]), simplifies the complexity of the problem. Additionally, the assumption is acceptable for slender beams. In such cases, this paper validates the assumption by comparing the model predictions with available experimental results in the literature.

The article is organized as follows. In Sect. 2, the problem is defined, and the nonlocal formulation and the solution technique are presented. The results obtained by the formulated model are presented and discussed in Sect. 3, which starts by comparing the predictions of the model with the experimental and theoretical results from the literature. Additionally, the section presents the results of the size-dependent frequency shifts of the beams with one, two, and three eccentric masses attached at different locations. The inverse problem is also investigated in Sect. 3. The concluding remarks are presented in Sect. 4.

## 2. Problem and Modeling

The problem is a mechanical mass sensor, illustrated in Fig. 1, which involves a miniaturized beam with a rectangular cross-section. The beam, made of a homogeneous isotropic material with Young's modulus $E$, has dimensions of length ($L$), in-plane thickness ($h$), and out-of-plane width ($b$). For the sake of generality, the formulation is developed for the case with the attachment of $n$ different point masses $M_i$ to the beam at various locations $x_i$, for $i = 1, \ldots, n$, with numbering proceeding from left to right. The distance between the mass centroid of the *i*-th particle and the mid-thickness of the beam is represented by $H_i$. Given that micro- and nanomechanical mass sensors typically adopt clamped-free or clamped-clamped configurations, only these boundary conditions are considered. The slender nature of the beam allows the utilization of the Bernoulli-Euler model to analyze the free transverse vibration under plane stress conditions. The problem is formulated using a Cartesian coordinate system, as depicted in Fig. 1.

It is possible to introduce a mass-spring system to model the connection of the particles to the mass sensor. The spring-like behavior may arise due to the local flexibility of the attachment point where the particle connects to the sensor. For instance, the spring-mass approach is used in [56] to account for the elastic connection between the nanoplate and the attached nanoparticle. The mass-spring representation of the attached particles forms a coupled mechanical system in which the motion of the sensor, spring, and attached particle are interconnected. While the spring-mass approach may offer a more general representation of real-world physics, it is notable that many researchers have developed analytical models under the assumption that the attached masses are perfectly connected to the sensor surface [41–45,47]. It is apparent from the data presented in the seminal papers [51,52,57] that the results obtained under this simplification are consistent with experimental findings. Similarly, we adopt the assumption here that point masses are perfectly connected to the sensor surface, implying they have no additional degree of freedom relative to the sensor. This assumption simplifies the model while still capturing essential aspects of sensor behavior.

To investigate the free transverse vibration of the micro- or nanobeam shown in Fig. 1, the domain is segmented at the mass locations. In the formulation, the superscript (*i*) on the left of a quantity indicates its association with the *i*-th segment, while on the right, it denotes the *i*-th derivative with respect to *x*. Additionally, a dot over a function signifies the derivative with respect to time, *t*.

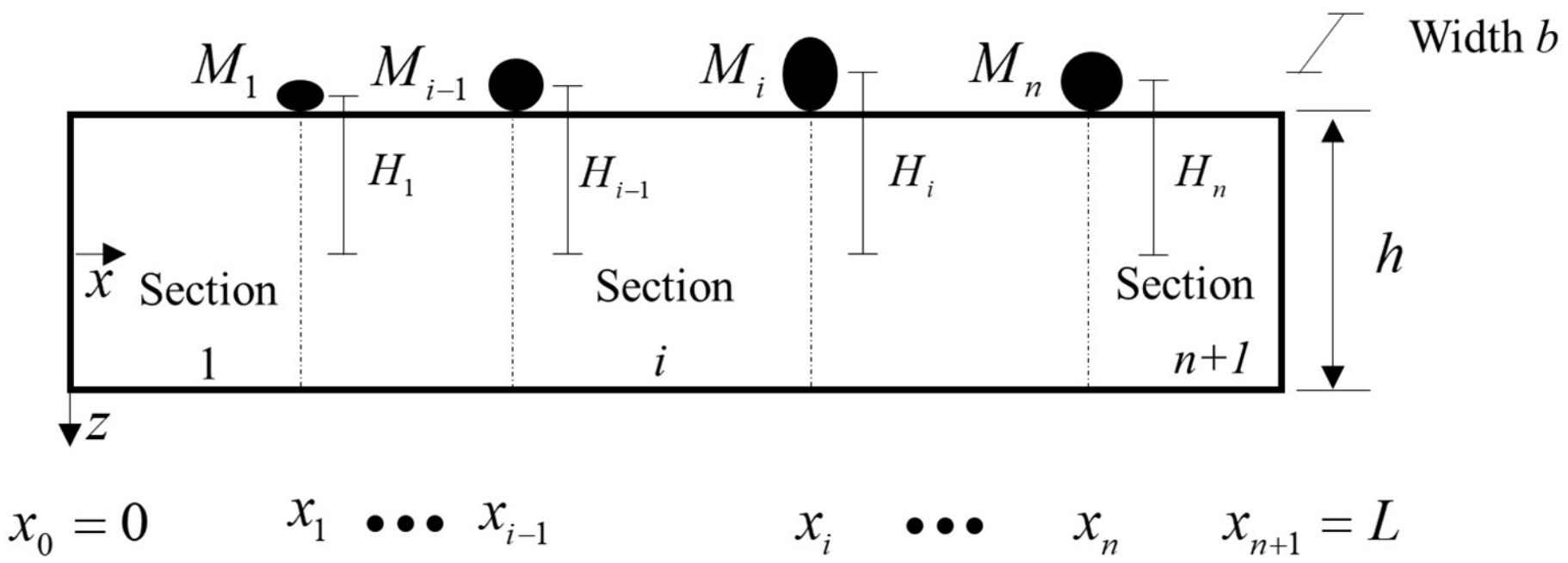

**Fig. 1** The miniaturized beam with attached particles.

### 2.1. Equations of Motion

Under the kinematic assumption of the Bernoulli-Euler theory, the displacement components of the *i*-th segment of the beam, where *i* ranges from 1 to *n* + 1, are expressed as follows:

$$
\begin{aligned}
{}^{(i)}u(x,z,t) &= {}^{(i)}w^{(1)}(x,t)\,z \\
{}^{(i)}v(x,z,t) &= {}^{(i)}w(x,t)
\end{aligned}
\tag{1}
$$

where ${}^{(i)}u(x,z,t)$ and ${}^{(i)}v(x,z,t)$ are, respectively, the axial and the transverse displacements in the *i*-th segment of the beam. The displacement function ${}^{(i)}w(x,t)$ refers to the transverse displacement of the mid-thickness. Using the linear displacement-strain relationship, the normal strain along the *x*-axis in the *i*-th segment of the beam, ${}^{(i)}\varepsilon(x,z,t)$, is derived as:

$$
{}^{(i)}\varepsilon(x,z,t) = {}^{(i)}w^{(2)}(x,t)\,z
\tag{2}
$$

for *i* = 1, …, *n* + 1.

The Hamilton principle is applied to obtain the weak form of the governing equations:

$$
\begin{aligned}
\int_{t_1}^{t_2} \Bigg\{ & \sum_{i=1}^{n+1} \int_A \int_{x_{i-1}}^{x_i} {}^{(i)}\sigma \delta^{(i)}\varepsilon dx dA + \sum_{i=1}^{n+1} \int_A \int_{x_{i-1}}^{x_i} \rho\, {}^{(i)}\ddot{u}\, \delta^{(i)}u dx dA + \sum_{i=1}^{n+1} \int_A \int_{x_{i-1}}^{x_i} \rho\, {}^{(i)}\ddot{v}\, \delta^{(i)}v dx dA + \\
& + \sum_{i=1}^{n} M_i\, {}^{(i)}\ddot{v}(x_i,H_i,t)\, \delta^{(i)}v(x_i,H_i,t) + \sum_{i=1}^{n} M_i\, {}^{(i)}\ddot{u}(x_i,H_i,t)\, \delta^{(i)}u(x_i,H_i,t) \Bigg\} dt = 0
\end{aligned}
\tag{3}
$$

Here, $t_1$ and $t_2$ represent arbitrary time instances, ${}^{(i)}\sigma$ and $\delta^{(i)}\varepsilon$ denote the axial stress and variation of the axial strain of the *i*-th segment of the beam, $\delta^{(i)}u$ and $\delta^{(i)}v$ are variations in displacements of the *i*-th segment of the beam, and $\rho$ is the mass density. The variations of displacements are assumed to be arbitrary, independent, and consistent with the compatibility conditions, vanishing at times $t_1$ and $t_2$. The first term on the left-hand side of Eq. (3) defines the contribution from the strain energy of the beam. The second and third terms pertain to the kinetic energy of the beam. The last two terms account for the kinetic energy of the attached particles associated with axial and transverse velocities.

By utilizing Eqs. (1) and (2) to express the variations in strain and displacements in terms of the displacement function ${}^{(i)}w(x,t)$ and applying Green's theorem wherever applicable, the following standard equations of motion are derived after some manipulations:

$$ {}^{(i)}M_{\text{moment}}^{(2)} - I_2 \, {}^{(i)}\ddot{w}^{(2)} + I_0 \, {}^{(i)}\ddot{w} = 0 \tag{4} $$

for $i$ = 1, …, $n$ + 1. The parameters $(I_o, I_2) = \int_A \rho(1, z^2) dA = (m, mh^2/12)$ where $m$ is the mass per unit length of the beam. The bending moment has a conventional definition of $M_{\text{moment}} = \int_A \sigma z dA$. The middle term on the left-hand side of the equation represents the effect of rotary inertia.

The equations of motions (4) are subjected to the following variationally consistent continuity conditions:

$$ \begin{aligned} &{}^{(i)}w(x_i,t) = {}^{(i+1)}w(x_i,t) \\ &{}^{(i)}w^{(1)}(x_i,t) = {}^{(i+1)}w^{(1)}(x_i,t) \\ &{}^{(i)}M_{\text{moment}}(x_i,t) + M_i \, H_i^2 \, {}^{(i)}\ddot{w}^{(1)}(x_i,t) = {}^{(i+1)}M_{\text{moment}}(x_i,t) \\ &{}^{(i)}M_{\text{moment}}^{(1)}(x_i,t) - I_2 {}^{(i)}\ddot{w}^{(1)}(x_i,t) - M_i {}^{(i)}\ddot{w}(x_i,t) = {}^{(i+1)}M_{\text{moment}}^{(1)}(x_i,t) - I_2 {}^{(i+1)}\ddot{w}^{(1)}(x_i,t) \end{aligned} \tag{5} $$

for $i$ = 1, …, $n$, and the boundary conditions at the beam's ends, $x = 0$ and $L$:

$$\left.\begin{aligned} {}^{(1)}w(x=0,t)=0 \\ {}^{(1)}w^{(1)}(x=0,t)=0 \end{aligned}\right\} \quad \text{clamped}$$

$$\left.\begin{aligned} {}^{(n+1)}w(x=L,t)=0 \\ {}^{(n+1)}w^{(1)}(x=L,t)=0 \end{aligned}\right\} \quad \text{if clamped} \tag{6}$$

$$\left.\begin{aligned} {}^{(n+1)}M_{\text{moment}}(x=L,t)=0 \\ {}^{(n+1)}M_{\text{moment}}^{(1)}(x=L,t)-I_2\,{}^{(n+1)}\ddot{w}^{(1)}(x=L,t)=0 \end{aligned}\right\} \quad \text{if free}$$

The scenario involving a micro- or nanobeam loaded with a single particle at the free end is described by Eqs. (4)-(6) for $x_1 = L$.

## 2.2. Size-Dependent Governing Equations in Terms of Displacements

In the stress-driven nonlocal theory, the constitutive equation of the beam, which establishes a relationship between curvature $\chi$ and moment $M_{\text{moment}}$, is expressed as [31]:

$$\chi(x,t)=\int_0^L \phi_{L_C}(x-\xi)\,CM_{\text{moment}}(\xi,t)\,d\xi \tag{7}$$

with $C = 1/(EI)$, where $I = bh^3/12$. The inclusion of size dependence in the formulation involves expressing curvature as the result of the integral convolution described above, considering moments across all cross-sections and the smoothing kernel function $\phi_{L_C}(x)$. The selection of the kernel function is quite flexible, as long as the chosen function is both mathematically and physically acceptable. For the sake of mathematical simplicity, the kernel function is commonly assumed as:

$$\phi_{L_C}(x)=\frac{1}{2L_C}e^{\left(-\frac{|x|}{L_C}\right)} \tag{8}$$

with $L_C$ being a characteristic length parameter. The particular kernel function given in Eq. (8) enables the conversion of the integral constitutive equation (7), which is over the entire beam length, to $n+1$ differential equations at various segments of the beam:

$$ {}^{(i)}\chi - L_C^2 \, {}^{(i)}\chi^{(2)} = C \, {}^{(i)}M_{\text{moment}} \tag{9} $$

for $i$ = 1, …, $n+1$, subjected to the following higher-order constitutive continuity conditions at the segment's ends:

$$ \begin{aligned} {}^{(i)}\chi^{(1)}\left(x_{i-1},t\right) &= \frac{1}{L_C}\left[ {}^{(i)}\chi\left(x_{i-1},t\right) - \sum_{k=1}^{i-1}\int_{x_{k-1}}^{x_k}\left(\frac{1}{L_C}e^{\frac{\xi - x_{i-1}}{L_C}} C \, {}^{(k)}M_{\text{moment}}\left(\xi,t\right)\right)d\xi \right] \\ {}^{(i)}\chi^{(1)}\left(x_{i},t\right) &= -\frac{1}{L_C}\left[ {}^{(i)}\chi\left(x_{i},t\right) - \sum_{k=i+1}^{n+1}\int_{x_{k-1}}^{x_k}\left(\frac{1}{L_C}e^{\frac{x_i - \xi}{L_C}} C \, {}^{(k)}M_{\text{moment}}\left(\xi,t\right)\right)d\xi \right] \end{aligned} \tag{10} $$

for $i$ = 1, …, $n+1$. The derivations of Eqs. (9) and (10) are purely mathematical and straightforward as presented in [36]. For $i=1$ at $x_0=0$ and $i=n+1$ at $x_{n+1}=L$, the constitutive continuity conditions (10) yield the constitutive boundary conditions [31]:

$$ \begin{aligned} {}^{(1)}\chi^{(1)}\left(0,t\right) &= \frac{1}{L_C} \, {}^{(1)}\chi\left(0,t\right) \\ {}^{(n+1)}\chi^{(1)}\left(L,t\right) &= -\frac{1}{L_C} \, {}^{(n+1)}\chi\left(L,t\right) \end{aligned} \tag{11} $$

Recalling the curvature-deflection relation of the Bernoulli-Euler theory, ${}^{(i)}\chi = {}^{(i)}w^{(2)}$, the constitutive equations (9)-(11) in terms of the transverse displacements, ${}^{(i)}w$, are:

$$ {}^{(i)}w^{(2)} - L_C^2 \, {}^{(i)}w^{(4)} = C \, {}^{(i)}M_{\text{moment}} \tag{12} $$

for $i$ = 1, …, $n+1$, and

$$
\begin{aligned}
{}^{(i)}w^{(3)}\left(x_{i-1},t\right) &= \frac{1}{L_C}\left[{}^{(i)}w^{(2)}\left(x_{i-1},t\right) - \sum_{k=1}^{i-1}\int_{x_{k-1}}^{x_k}\left(\frac{1}{L_C}e^{\frac{\xi - x_{i-1}}{L_C}}\left[{}^{(k)}w^{(2)}\left(\xi,t\right) - L_C^2\,{}^{(k)}w^{(4)}\left(\xi,t\right)\right]\right)d\xi\right] \\
{}^{(i)}w^{(3)}\left(x_i,t\right) &= -\frac{1}{L_C}\left[{}^{(i)}w^{(2)}\left(x_i,t\right) - \sum_{k=i+1}^{n+1}\int_{x_{k-1}}^{x_k}\left(\frac{1}{L_C}e^{\frac{x_i - \xi}{L_C}}\left[{}^{(k)}w^{(2)}\left(\xi,t\right) - L_C^2\,{}^{(k)}w^{(4)}\left(\xi,t\right)\right]\right)d\xi\right]
\end{aligned}
\tag{13}
$$

for $i = 1, \ldots, n+1$, which for $i = 1$ at $x_0 = 0$ and $i = n+1$ at $x_{n+1} = L$ yield

$$
\begin{aligned}
{}^{(1)}w^{(3)}\left(0,t\right) &= \frac{1}{L_C}\,{}^{(1)}w^{(2)}\left(0,t\right) \\
{}^{(n+1)}w^{(3)}\left(L,t\right) &= -\frac{1}{L_C}\,{}^{(n+1)}w^{(2)}\left(L,t\right)
\end{aligned}
\tag{14}
$$

Now, the governing Eqs. (4)-(6) can be written solely in terms of the transverse displacements, ${}^{(i)}w$, using the bending moment definition given by the constitutive equation (12):

$$
L_C^2\,{}^{(i)}w^{(6)} - {}^{(i)}w^{(4)} + \frac{Cmh^2}{12}\,{}^{(i)}\ddot{w}^{(2)} - Cm\,{}^{(i)}\ddot{w} = 0 \tag{15}
$$

for $i = 1, \ldots, n+1$, and

$$
\begin{aligned}
&{}^{(i)}w\left(x_i,t\right) = {}^{(i+1)}w\left(x_i,t\right) \\
&{}^{(i)}w^{(1)}\left(x_i,t\right) = {}^{(i+1)}w^{(1)}\left(x_i,t\right) \\
&L_C^2\,{}^{(i)}w^{(4)}\left(x_i,t\right) - {}^{(i)}w^{(2)}\left(x_i,t\right) - CM_i\,H_i^2\,{}^{(i)}\ddot{w}^{(1)}\left(x_i,t\right) = L_C^2\,{}^{(i+1)}w^{(4)}\left(x_i,t\right) - {}^{(i+1)}w^{(2)}\left(x_i,t\right) \\
&L_C^2\,{}^{(i)}w^{(5)}\left(x_i,t\right) - {}^{(i)}w^{(3)}\left(x_i,t\right) + \frac{Cmh^2}{12}\,{}^{(i)}\ddot{w}^{(1)}\left(x_i,t\right) + CM_i\,{}^{(i)}\ddot{w}\left(x_i,t\right) = \\
&\qquad L_C^2\,{}^{(i+1)}w^{(5)}\left(x_i,t\right) - {}^{(i+1)}w^{(3)}\left(x_i,t\right) + \frac{Cmh^2}{12}\,{}^{(i+1)}\ddot{w}^{(1)}\left(x_i,t\right)
\end{aligned}
\tag{16}
$$

for $i = 1, \ldots, n$, and

$$\left.\begin{aligned} {}^{(1)}w\left(x=0,t\right)=0 \\ {}^{(1)}w^{(1)}\left(x=0,t\right)=0 \end{aligned}\right\} \quad \text{clamped}$$

$$\left.\begin{aligned} {}^{(n+1)}w\left(x=L,t\right)=0 \\ {}^{(n+1)}w^{(1)}\left(x=L,t\right)=0 \end{aligned}\right\} \quad \text{if clamped} \tag{17}$$

$$\left.\begin{aligned} & L_C^2\,{}^{(n+1)}w^{(4)}\left(x=L,t\right)-{}^{(n+1)}w^{(2)}\left(x=L,t\right)=0 \\ & L_C^2\,{}^{(n+1)}w^{(5)}\left(x=L,t\right)-{}^{(n+1)}w^{(3)}\left(x=L,t\right)+\frac{Cmh^2}{12}\,{}^{(n+1)}\ddot{w}^{(1)}\left(x=L,t\right)=0 \end{aligned}\right\} \quad \text{if free}$$

The dynamic equilibrium equation (15), along with the variationally consistent continuity and boundary conditions (16) and (17), as well as the constitutive continuity and boundary conditions (13) and (14), all expressed solely in terms of the transverse displacements of the beam segments, ${}^{(i)}w$, define the size-dependent free transverse vibration of the loaded beam. In the next section, the equation of motion (15) will be solved, and dimensionless equations will be provided for the calculation of the natural frequencies.

### 2.3. Natural Frequency Determination

The analytical technique of separating spatial and time variables is utilized to solve the equation of motion (15). Therefore, the following form of the solution is assumed: ${}^{(i)}w\left(x,t\right)={}^{(i)}\psi\left(x\right)e^{i\omega t}$. The parameters ${}^{(i)}\psi$ and $\omega$ are, respectively, the spatial mode shape of section $i$ and the natural frequency of free vibrations of the beam. Substitution of the assumed form of the solution into the equation of motion yields the spatial differential equation, which is represented in a dimensionless form using the following dimensionless parameters:

$${}^{(i)}\bar{\psi}=\frac{{}^{(i)}\psi}{L};\ \ \bar{x}=\frac{x}{L};\ \ \bar{h}=\frac{h}{L};\ \ \bar{\omega}=\omega L^2\sqrt{Cm};\ \ \bar{H}_i=\frac{H_i}{L};\ \ \bar{M}_i=\frac{M_i}{mL};\ \ \lambda=\frac{L_C}{L} \tag{18}$$

as:

$$\lambda^2\, {}^{(i)}\overline{\psi}^{(6)}\left(\overline{x}\right) - {}^{(i)}\overline{\psi}^{(4)}\left(\overline{x}\right) + \overline{\omega}^2\left[{}^{(i)}\overline{\psi}\left(\overline{x}\right) - \frac{\overline{h}^2}{12}\, {}^{(i)}\overline{\psi}^{(2)}\left(\overline{x}\right)\right] = 0 \tag{19}$$

for $i = 1, \ldots, n+1$. The dimensionless variationally consistent continuity and boundary conditions are:

$$\begin{aligned}
&{}^{(i)}\overline{\psi}\left(\overline{x}_i\right) = {}^{(i+1)}\overline{\psi}\left(\overline{x}_i\right) \\
&{}^{(i)}\overline{\psi}^{(1)}\left(\overline{x}_i\right) = {}^{(i+1)}\overline{\psi}^{(1)}\left(\overline{x}_i\right) \\
&\lambda^2\, {}^{(i)}\overline{\psi}^{(4)}\left(\overline{x}_i\right) - {}^{(i)}\overline{\psi}^{(2)}\left(\overline{x}_i\right) + \overline{M}_i\, \overline{\omega}^2\, \overline{H}_i^2\, {}^{(i)}\overline{\psi}^{(1)}\left(\overline{x}_i\right) = \lambda^2\, {}^{(i+1)}\overline{\psi}^{(4)}\left(\overline{x}_i\right) - {}^{(i+1)}\overline{\psi}^{(2)}\left(\overline{x}_i\right) \\
&\lambda^2\, {}^{(i)}\overline{\psi}^{(5)}\left(\overline{x}_i\right) - {}^{(i)}\overline{\psi}^{(3)}\left(\overline{x}_i\right) - \frac{\overline{\omega}^2\overline{h}^2}{12}\, {}^{(i)}\overline{\psi}^{(1)}\left(\overline{x}_i\right) - \overline{M}_i\overline{\omega}^2\overline{\psi}\left(\overline{x}_i\right) = \\
&\qquad\qquad \lambda^2\, {}^{(i+1)}\overline{\psi}^{(5)}\left(\overline{x}_i\right) - {}^{(i+1)}\overline{\psi}^{(3)}\left(\overline{x}_i\right) - \frac{\overline{\omega}^2\overline{h}^2}{12}\, {}^{(i+1)}\overline{\psi}^{(1)}\left(\overline{x}_i\right)
\end{aligned} \tag{20}$$

for $i = 1, \ldots, n$, and

$$\left.\begin{aligned}
&{}^{(1)}\overline{\psi}\left(0\right) = 0 \\
&{}^{(1)}\overline{\psi}^{(1)}\left(0\right) = 0
\end{aligned}\right\} \quad \text{clamped}$$

$$\left.\begin{aligned}
&{}^{(n+1)}\overline{\psi}\left(1\right) = 0 \\
&{}^{(n+1)}\overline{\psi}^{(1)}\left(1\right) = 0
\end{aligned}\right\} \quad \text{if clamped} \tag{21}$$

$$\left.\begin{aligned}
&\lambda^2\, {}^{(n+1)}\overline{\psi}^{(4)}\left(1\right) - {}^{(n+1)}\overline{\psi}^{(2)}\left(1\right) = 0 \\
&\lambda^2\, {}^{(n+1)}\overline{\psi}^{(5)}\left(1\right) - {}^{(n+1)}\overline{\psi}^{(3)}\left(1\right) - \frac{\overline{\omega}^2\overline{h}^2}{12}\, {}^{(n+1)}\overline{\psi}^{(1)}\left(1\right) = 0
\end{aligned}\right\} \quad \text{if free}$$

Additionally, the dimensionless constitutive continuity and boundary conditions are:

$$
\begin{aligned}
{}^{(i)}\overline{\psi}^{(3)}\left(\overline{x}_{i-1}\right) &= \frac{1}{\lambda}\left[{}^{(i)}\overline{\psi}^{(2)}\left(\overline{x}_{i-1}\right) - \sum_{k=1}^{i-1}\int_{\overline{x}_{k-1}}^{\overline{x}_k}\left(\frac{1}{\lambda}e^{\frac{\xi-\overline{x}_{i-1}}{\lambda_c}}\left[{}^{(k)}\overline{\psi}^{(2)}\left(\xi\right) - \lambda^2\,{}^{(k)}\overline{\psi}^{(4)}\left(\xi\right)\right]\right)d\xi\right] \\
{}^{(i)}\overline{\psi}^{(3)}\left(\overline{x}_{i}\right) &= -\frac{1}{\lambda}\left[{}^{(i)}\overline{\psi}^{(2)}\left(\overline{x}_{i}\right) - \sum_{k=i+1}^{n+1}\int_{\overline{x}_{k-1}}^{\overline{x}_k}\left(\frac{1}{\lambda}e^{\frac{\overline{x}_{i}-\xi}{\lambda_c}}\left[{}^{(k)}\overline{\psi}^{(2)}\left(\xi\right) - \lambda^2\,{}^{(k)}\overline{\psi}^{(4)}\left(\xi\right)\right]\right)d\xi\right]
\end{aligned} \tag{22}
$$

for $i = 1, \ldots, n+1$, which for $i=1$ at $\overline{x}_0 = 0$ and $i = n+1$ at $\overline{x}_{n+1} = 1$ yield,

$$
\begin{aligned}
{}^{(1)}\overline{\psi}^{(3)}\left(0\right) &= \frac{1}{\lambda}\,{}^{(1)}\overline{\psi}^{(2)}\left(0\right) \\
{}^{(n+1)}\overline{\psi}^{(3)}\left(1\right) &= -\frac{1}{\lambda}\,{}^{(n+1)}\overline{\psi}^{(2)}\left(1\right)
\end{aligned} \tag{23}
$$

The spatial equation (19) for the *i*-th segment of the beam is a linear, homogeneous, sixth-order ordinary differential equation with constant coefficients. This equation is solved in a closed form utilizing the solution technique described next. The general form of the solution ${}^{(i)}\overline{\psi} = {}^{(i)}ae^{{}^{(i)}\beta\overline{x}}$ with ${}^{(i)}a$ and ${}^{(i)}\beta$ being unknown constants, is substituted into Eq. (19), to obtain the corresponding sixth-order algebraic characteristic equations in terms of ${}^{(i)}\beta$. Using the change of variable technique, the characteristic equation is further simplified to a third-order algebraic equation, for which the closed-form solutions are available in the mathematics handbooks. Therefore, the solution of Eq. (19) is defined as ${}^{(i)}\overline{\psi} = \sum_{j=1}^{6}{}^{(i)}a_j e^{{}^{(i)}\beta_j\overline{x}}$ in terms of six unknown constants ${}^{(i)}a_j$ for $j = 1, \ldots, 6$, which makes a total of $6\times\left(n+1\right)$ unknown constants for the beam with $n$ attached particles. The solutions must satisfy $4\times n$ variationally consistent continuity conditions in Eq. (20), 4 variationally consistent boundary conditions in Eq. (21), and $2\times\left(n+1\right)$ constitutive continuity and boundary conditions in Eqs. (22) and (23). This results in a homogeneous system of $6\times\left(n+1\right)$ algebraic equations for $6\times\left(n+1\right)$ unknown constants. The non-trivial solution exists only if the determinant of the coefficient matrix vanishes, which leads to the determination of the natural frequencies of the micro- or nanobeam with *n* attached particles. To determine the natural frequencies, the bisection method is employed to find the roots of the determinant of the coefficient matrix.

Note that the eccentricity of the attached masses, $\bar{H}_i$ for $i = 1, \ldots, n$, is incorporated in Eq. (20). Setting their values to zero in this equation results in the current formulation reducing to that presented in [38].

## 3. Results and Discussion

In this section, we analyze cantilever and clamped-clamped miniaturized beams with one to three attached particles.

### 3.1. Verification

The validity of the formulated model is verified for five different cases against the experimental and numerical results available in the literature. The predictions of the model with $\lambda = 0$ (local model) and the experimental data reported in [58] are compared in Fig. 2. The results refer to the frequency shifts of the first four modes of vibration in a microcantilever sensor made of $SiO_2$ with a thickness of 940 nm coated with 10 nm Ti and 100 nm Au, after the attachment of a single gold bead with a radius of 0.9 μm and a mass of 60 pg at different locations. The length, thickness, and width of the microcantilever are, respectively, 153, 1.05 and 11 microns. To calculate the distance between the mass center of the gold particle and the mid-thickness of the sensor, $H_1$, the gold particle is assumed to be spherical. The Young's modulus and the density of $SiO_2$, Ti, and Au, are, respectively, 70, 110, and 57 GPa, and 2150, 4500, and 19300 kg/m$^3$ [58]. Since the Ti layer with the highest Young's modulus is relatively thin, and the Young's modulus of the $SiO_2$ and Au layers are similar, the multilayered sensor is assumed to be homogeneous with the elastic modulus equal to 69 GPa calculated by the rule of mixture. The effective density of the sensor is also calculated as 3805 kg/m$^3$ using the rule of mixture. As can be seen in Fig. 2, the model can well predict the experimentally measured percentage change in frequency due to the attachment of the gold particle at different locations across the sensor for the nonlocal parameter equal to zero. This confirms the assumption made in the development of the model regarding the decoupling of bending and axial vibrations in slender beams with eccentric masses.

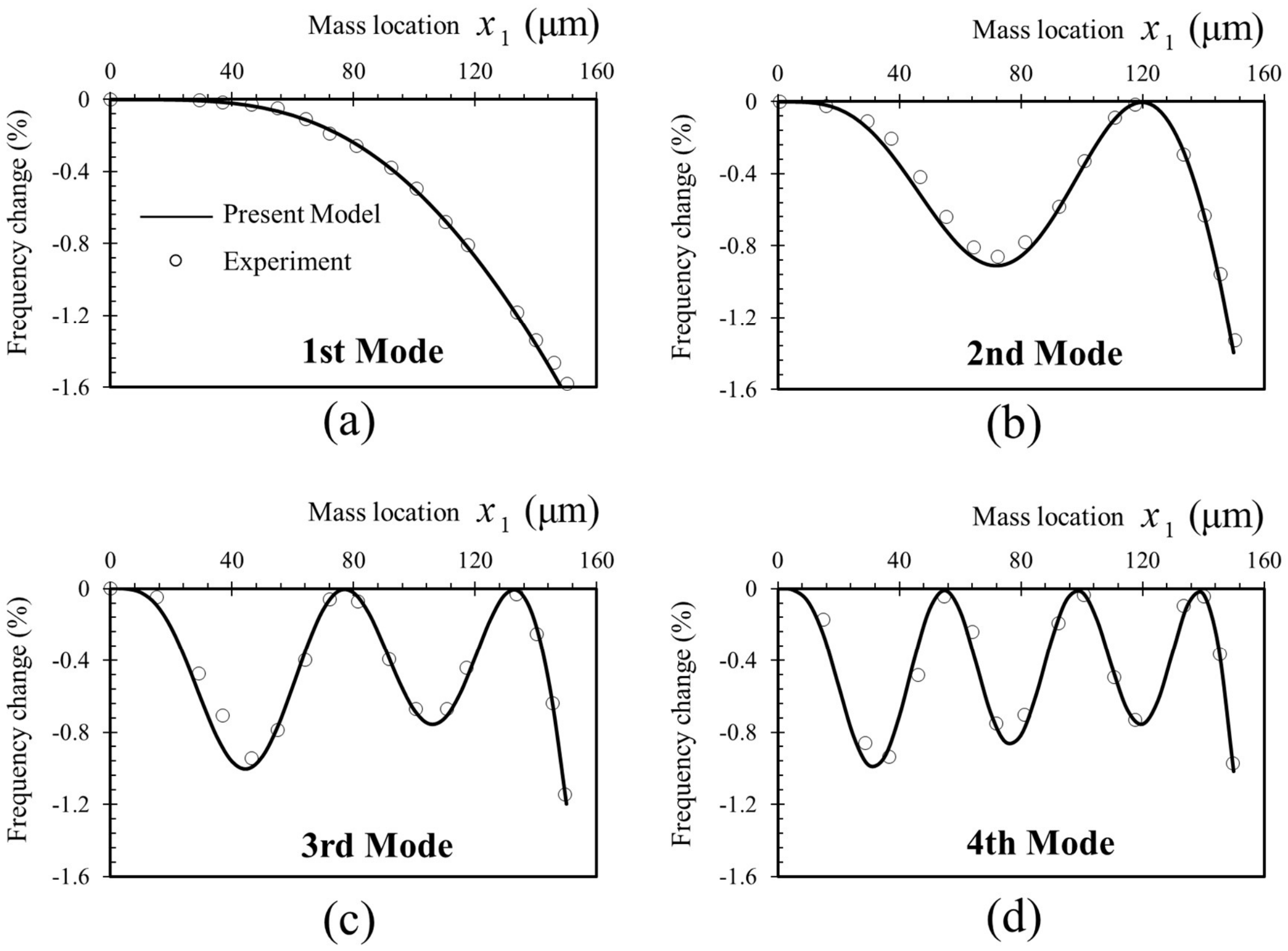

**Fig. 2** Comparisons between the predictions of the present model and the experimental results reported in [58] for the frequency reduction of the (a) first, (b) second, (c) third, and (d) fourth modes of vibration.

In [49], frequencies of nanocantilevers with a single attached particle positioned at the free end are obtained using the stress-driven nonlocal theory and Bernoulli-Euler beam model. In the formulation, the eccentricity of the tip mass is neglected. The first and second frequencies predicted by the current model (dashed lines) and those presented in [49] (circle markers) for the case of a nonlocal nanocantilever with $\lambda = 0.04$ and $\overline{h} = 0.0385$ are depicted in Fig. 3 by varying the tip mass. As can be seen in the figure, the results of the present model virtually coincide with those presented in [49].

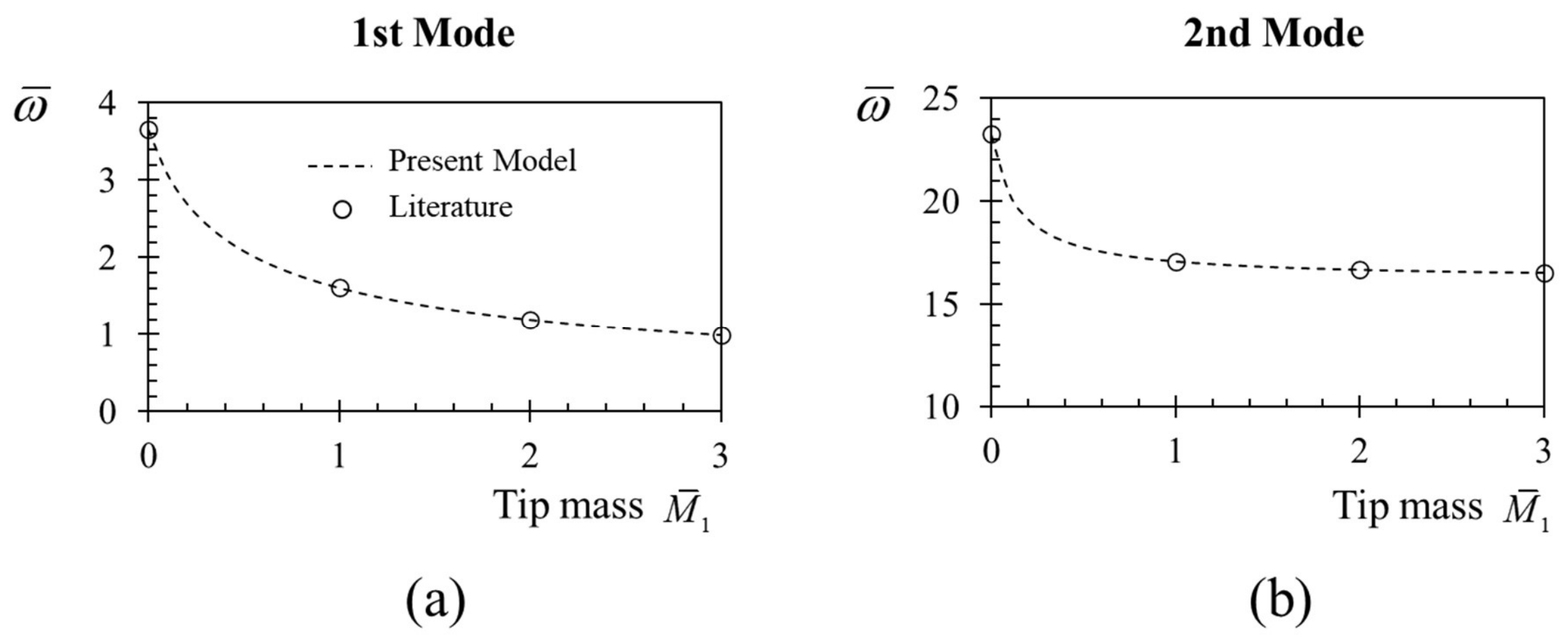

**Fig. 3** Dimensionless natural frequencies $\bar{\omega}$ of a cantilever beam with an attached mass at the free end for $\lambda = 0.04$ and $\bar{h} = 0.0385$. The eccentricity of the tip mass is neglected. Results are presented for the first and second modes of vibration, varying the tip mass $\bar{M}_1$, and compared with those obtained in [49] using the stress-driven nonlocal theory and Bernoulli-Euler beam model.

The dimensionless natural frequencies $\bar{\omega}$ of the first three modes of vibration of a cantilever beam with an attached mass at the free end, $\bar{x}_1 = 1$, with $\lambda = 0$ (local model), and $H_1/h = 2$ are presented in Table 1 for different dimensionless thicknesses and masses. The results are in good agreement with those obtained in [47] through the Generalized Differential/Integral Quadrature method based on the Bernoulli-Euler beam theory and considering coupling between the axial and transverse vibrations due to the eccentricity of the attached mass. The reduction in the natural frequencies is higher when the beam is loaded by a heavier particle. The reduction in the natural frequency is also higher at higher modes of vibration. For instance for the cantilever beam with $\bar{h} = 0.1$, changing the mass of the attached particle at the free end from $\bar{M}_1 = 0.01$ to 0.2 causes 25, 32, and 39% reduction in the natural frequency of the, respectively, first, second, and third mode of vibration.

**Table 1** Dimensionless natural frequencies $\bar{\omega}$ of a cantilever beam with an attached mass at the free end, $\bar{x}_1 = 1$, for $\lambda = 0$ (local model), and $H_1/h = 2$. Results are presented for the first three modes of vibrations, different values of $\bar{h}$ and $\bar{M}_1$, and compared with the numerical results obtained in [47] through the Generalized Differential/Integral Quadrature method based on the Bernoulli-Euler beam theory and considering coupling between the axial and transverse vibrations due to the eccentricity of the attached mass.

| $\bar{h} = 0.05$ | | | | | | |
|---|---|---|---|---|---|---|
| *Mode* | $\bar{M}_1 = 0.01$ | | $\bar{M}_1 = 0.1$ | | $\bar{M}_1 = 0.2$ | |
| | Present Model | Ref. [47] | Present Model | Ref. [47] | Present Model | Ref. [47] |
| 1st | 3.4448 | 3.4426 | 2.9582 | 2.9564 | 2. 6000 | 2.5983 |
| 2nd | 21.4554 | 21.4550 | 18.6331 | 18.6285 | 17.1033 | 17.0906 |
| 3rd | 59.3869 | 59.3840 | 49.9877 | 49.6582 | 45.0700 | 44.3382 |
| $\bar{h} = 0.1$ | | | | | | |
| *Mode* | $\bar{M}_1 = 0.01$ | | $\bar{M}_1 = 0.1$ | | $\bar{M}_1 = 0.2$ | |
| | Present Model | Ref. [47] | Present Model | Ref. [47] | Present Model | Ref. [47] |
| 1st | 3.4362 | 3.4341 | 2.9297 | 2.9278 | 2.5621 | 2.5604 |
| 2nd | 20.9716 | 20.9715 | 16.6354 | 16.5804 | 14.3308 | 14.2235 |
| 3rd | 55.9357 | 53.3060 | 39.2107 | 37.4013 | 33.9692 | 32.2436 |

The dimensionless natural frequencies $\bar{\omega}$ of a cantilever beam with an attached mass at the free end, $\bar{x}_1 = 1$, as well as at the mid-span, $\bar{x}_1 = 0.5$, for $\lambda = 0$ (the local model), $\bar{h} = 0.01$, and $\bar{H}_1 = 0.02$ are presented in Table 2. Results are presented for the first two modes of vibrations, and the dimensionless mass equal to $\bar{M}_1 = 0.005$, 0.02, and 0.04. The predictions of the model are compared with the numerical results obtained in [47] through the finite element method. The present model can predict the natural frequencies of the cantilever beam with excellent accuracy in comparison with the numerical results. It can be understood from the results in Table 2 that the effect of the attached particle on the natural frequencies of both modes reduces when the particle is located at the mid-span. For the dimensionless mass equal to $\bar{M}_1 = 0.04$, the natural frequencies of the first and second modes of vibration increase by 7 and 3% when the location of the attached particle changes from the free end to the mid-span.

**Table 2** Dimensionless natural frequencies $\bar{\omega}$ of a cantilever beam with an attached mass at the free end, $\bar{x}_1 = 1$, and at the mid-span, $\bar{x}_1 = 0.5$, for $\lambda = 0$ (local model), $\bar{h} = 0.01$, and $\bar{H}_1 = 0.02$. Results are presented for the first two modes of vibrations, and different values of $\overline{M}_1$, and compared with the numerical results obtained in [47] through the FE Method.

| $\bar{x}_1 = 1$ | | | | | | |
|---|---|---|---|---|---|---|
| *Mode* | $\overline{M}_1 = 0.005$ | | $\overline{M}_1 = 0.02$ | | $\overline{M}_1 = 0.04$ | |
| | Present Model | Ref. [47] | Present Model | Ref. [47] | Present Model | Ref. [47] |
| 1st | 3.4813 | 3.4814 | 3.3828 | 3.3830 | 3.2634 | 3.2636 |
| 2nd | 21.8161 | 21.8209 | 21.2423 | 21.2522 | 20.6141 | 20.6299 |

| $\bar{x}_1 = 0.5$ | | | | | | |
|---|---|---|---|---|---|---|
| *Mode* | $\overline{M}_1 = 0.005$ | | $\overline{M}_1 = 0.02$ | | $\overline{M}_1 = 0.04$ | |
| | Present Model | Ref. [47] | Present Model | Ref. [47] | Present Model | Ref. [47] |
| 1st | 3.5119 | 3.5120 | 3.4998 | 3.4999 | 3.4838 | 3.4840 |
| 2nd | 21.9204 | 21.9233 | 21.5993 | 21.6021 | 21.1977 | 21.2005 |

The formulation presented in this paper can be readily applied to calculate the natural frequencies of clamped-clamped beams with multiple particle attachments. This is shown in Table 3, where the dimensionless natural frequencies $\bar{\omega}$ of the first three modes of vibration of a clamped-clamped beam with $\lambda = 0$ (local model), $\bar{h} = 0.01$, and loaded by two masses having $\overline{H}_1 = \overline{H}_2 = 0.02$ are presented on varying the dimensionless mass and location of the attached particles. The numerical results obtained in [59] through the Laplace transform are also presented. The predictions of the model are in good agreement with the numerical results. The difference between the results is because the numerical results were obtained neglecting the rotary inertia term and the distance between the mass centroid of the particles and the mid-thickness of the beam. As can be seen in Table 3, the predictions of the model are identical to the numerical data for the case with $\bar{h} = \overline{H} = 0$, which eliminates the effects of the rotary inertia and distance between the mass centroid of the particles and the mid-thickness of the beam from the formulation.

**Table 3** Dimensionless natural frequencies $\bar{\omega}$ of a clamped-clamped beam with two attached masses, $\lambda = 0$ (local model), $\bar{h} = 0.01$, and $\overline{H}_1 = \overline{H}_2 = 0.02$. Results are presented for the first three modes of vibrations,

different values of masses and their locations, and compared with the numerical data obtained in [59] through the Laplace transform.

| $\bar{x}_1 = 0.25$; $\bar{x}_2 = 0.75$ | | | | | | |
|---|---|---|---|---|---|---|
| *Mode* | $\bar{M}_1 = \bar{M}_2 = 0.01$ | | | $\bar{M}_1 = \bar{M}_2 = 0.1$ | | |
| | Present Model | Present Model $\bar{h} = \bar{H} = 0$ | Ref. [59] | Present Model | Present Model $\bar{h} = \bar{H} = 0$ | Ref. [59] |
| 1st | 22.2055 | 22.2082 | 22.2082 | 20.8381 | 20.8552 | 20.8552 |
| 2nd | 60.4102 | 60.4223 | 60.4223 | 51.6927 | 51.7112 | 51.7112 |
| 3rd | 118.6536 | 118.7195 | 118.7195 | 104.5430 | 104.7477 | 104.7477 |
| $\bar{x}_1 = 1/3$; $\bar{x}_2 = 2/3$ | | | | | | |
| *Mode* | $\bar{M}_1 = \bar{M}_2 = 0.01$ | | | $\bar{M}_1 = \bar{M}_2 = 0.1$ | | |
| | Present Model | Present Model $\bar{h} = \bar{H} = 0$ | Ref. [59] | Present Model | Present Model $\bar{h} = \bar{H} = 0$ | Ref. [59] |
| 1st | 22.0373 | 22.0392 | 22.0392 | 19.5667 | 19.5765 | 19.5765 |
| 2nd | 60.4265 | 60.4361 | 60.4361 | 51.7571 | 51.7786 | 51.7786 |
| 3rd | 120.5615 | 120.7219 | 120.7219 | 118.2381 | 119.4187 | 119.4187 |

### 3.2. Eccentricity Effect

The effect of the eccentricity of the attached mass on the first four natural frequencies of both cantilever and clamped-clamped beams is investigated in this section. Two different scenarios are considered. Firstly, Fig. 4 and Fig. 5 depict the frequencies as they vary by changing the eccentricity and location of an attached particle, while maintaining constant the thickness-to-length ratio of the beam and the intensity of the mass, $\bar{h} = \bar{M}_1 = 0.1$. The results are presented for the nonlocal beam with the dimensionless nonlocal parameter set to 0.3.

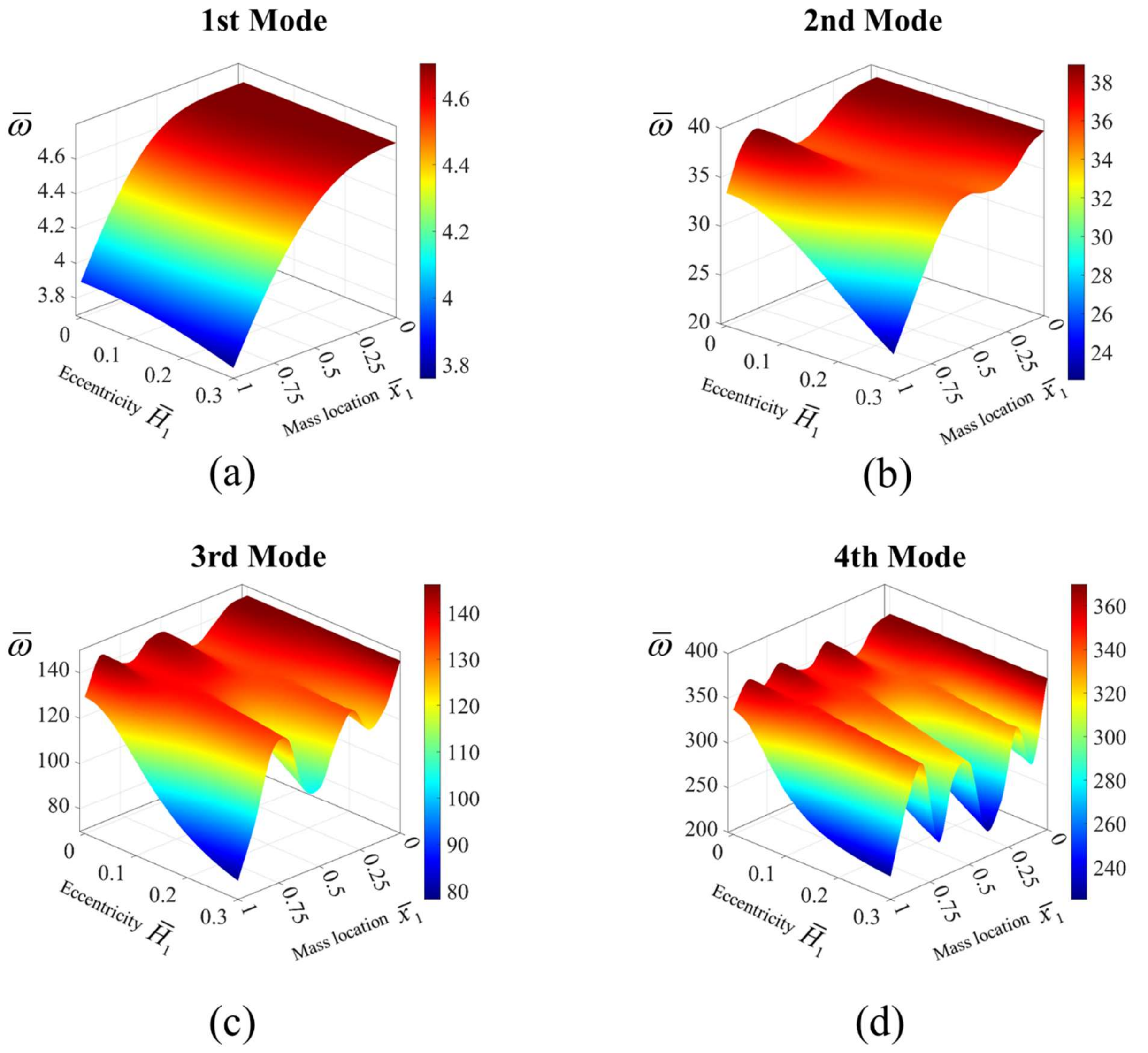

**Fig. 4** Frequencies vs. mass location and its eccentricity for the first four vibration modes in a miniaturized cantilever sensor with $\bar{h} = \bar{M}_1 = 0.1$ and $\lambda = 0.3$ .

The impact of eccentricity on the natural frequencies varies depending on the location where the particle is attached to the beam. For example, in Fig. 3(a) depicting the first mode of vibration for a cantilever beam, the influence of eccentricity on the frequency is more pronounced when the particle is attached closer to the free end. This enhanced effect stems from the relationship between the eccentricity and the axial velocity of the attached particle; thus, as the particle exhibits higher axial velocities, the effect becomes more pronounced. Referring to the displacement field provided in Eq.(1), higher axial velocities are associated with increased bending slopes. In the case of the first mode of vibration for a cantilever beam, this phenomenon occurs at locations closer to the free end. Consequently, the effect of eccentricity on the frequencies is expected to diminish when the particle is attached at locations with lower bending slopes. This reasoning explains why the effect of eccentricity on the second frequency is minimized when the particle is attached at $\bar{x}_1 = 0.5$ and 0.6, as depicted in Fig. 4(b). Similarly, since the third and fourth mode shapes feature two and three local maxima, respectively, where the bending slope vanishes, the surfaces in Fig. 4(c) and Fig. 4(d) exhibit two and three peaks, respectively, where the frequencies remain independent of the eccentricity of the attached mass.

The first four frequencies of a clamped-clamped nonlocal beam with $\lambda = 0.3$ and $\bar{h} = \bar{M}_1 = 0.1$ are illustrated in Fig. 5. These frequencies are displayed while varying the eccentricity and location of the attached mass. Similar to the behavior observed for the cantilever beam in Fig. 4, the impact of eccentricity on the frequencies is primarily controlled by the location of the attached mass. All the presented surfaces exhibit symmetry with respect to the plane $\bar{x}_1 = 0.5$ due to the inherent symmetry of the clamped-clamped beam. In particular, the first and third frequencies remain unaffected by eccentricity when the mass is attached at the mid-span, where the bending slope of the beam remains consistently zero throughout the vibration. Generally, for each mode of vibration from first to fourth, there exist one to four specific locations where the eccentricity of the attached mass does not influence the frequencies due to the vanishing bending slopes at these locations. Furthermore, in both the cantilever and clamped-clamped beams depicted in Fig. 4 and Fig. 5, the impact of the mass location on the natural frequencies intensifies with increasing eccentricity of the attached particles. As the eccentricity increases, the attached particle deviates further from the mid-thickness of the sensor, leading to a higher axial velocity and consequently, higher contribution to the kinetic energy of the system. In this scenario, the presence of the attached particle becomes more influential in the dynamics of the system, resulting in any change in its position having a greater impact on the natural frequencies.

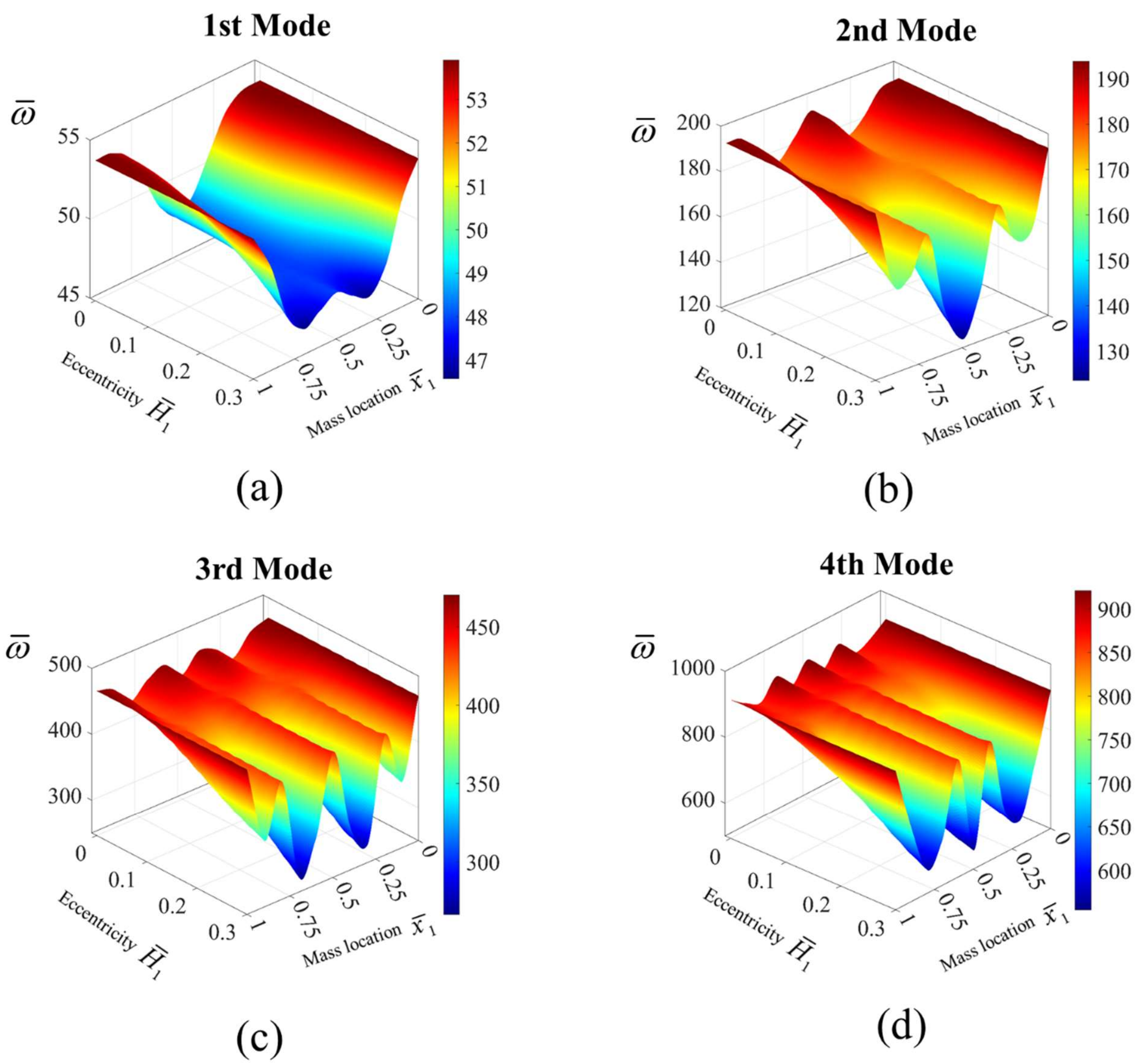

**Fig. 5** Frequencies vs. mass location and its eccentricity for the first four vibration modes in a miniaturized clamped-clamped sensor with $\bar{h} = \bar{M}_1 = 0.1$ and $\lambda = 0.3$.

The first four frequencies of a cantilever nonlocal beam with $\lambda = 0.3$ and an attached particle with $\bar{M}_1 = 0.1$ positioned at the free end, $\bar{x}_1 = 1$, are depicted in Fig. 6. These frequencies are shown while varying the eccentricity of the attached particle and the aspect ratio of the beam, i.e., the thickness-to-length ratio. Sensors with higher aspect ratios experience a greater effect of rotary inertia, resulting in increased dynamic loading on the sensor due to the additional inertia loading caused by the rotational acceleration of its elements. Therefore, as shown in Fig. 6, sensors with higher aspect ratios exhibit lower frequencies for all modes of vibration. The reduction in frequencies due to higher aspect ratios is particularly significant when analyzing vibration modes at higher frequencies. This is because the vibrating configurations of the sensors associated with the higher modes of vibration are divided into relatively short segments by nodal points, making the effect of rotary inertia more significant. Additionally, it is evident from the figure that the impact of the eccentricity of the attached mass on the frequencies is influenced by the aspect ratio.

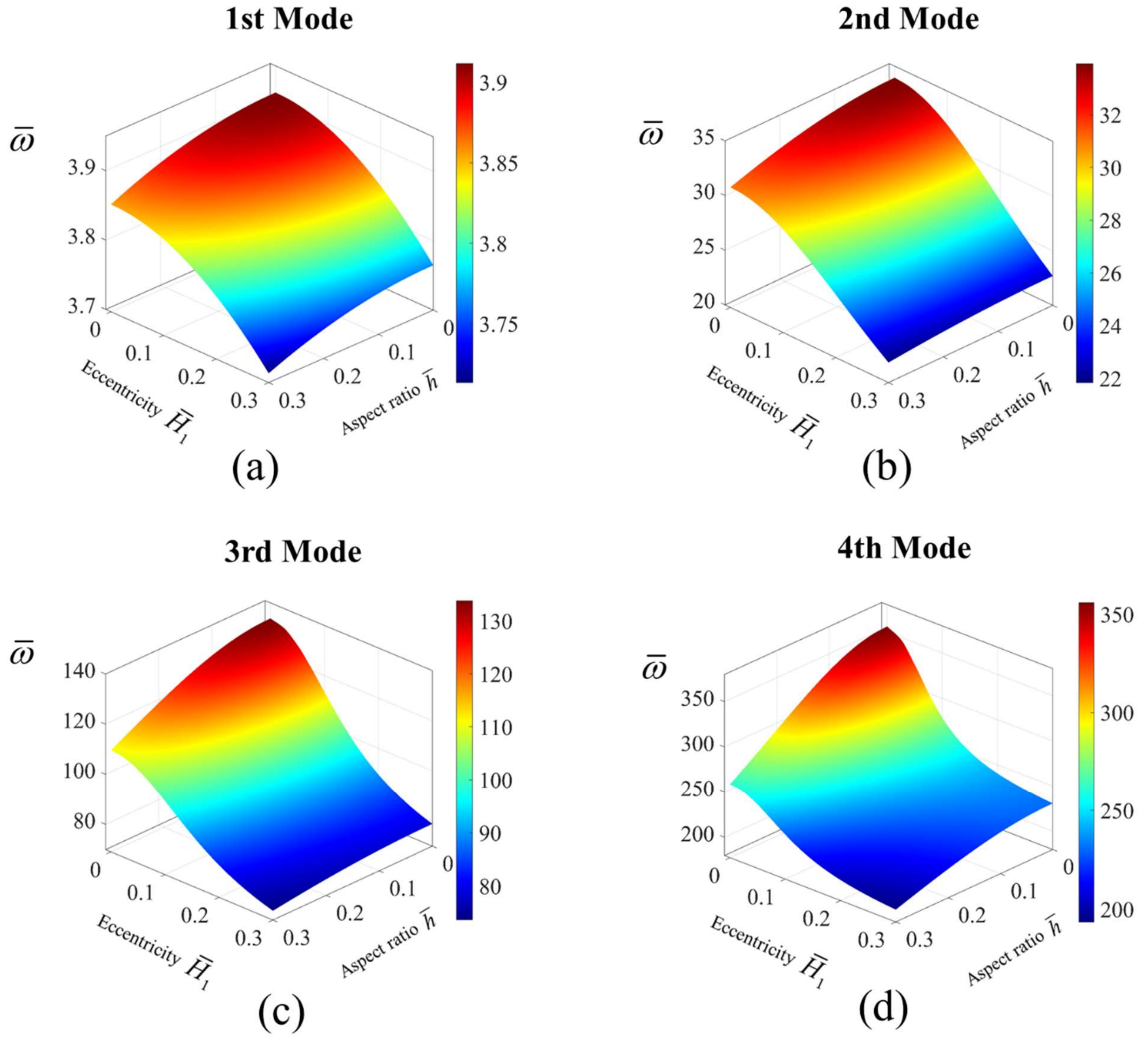

**Fig. 6** Frequencies vs. mass eccentricity and aspect ratio of the miniaturized cantilever sensor with $\bar{M}_1 = 0.1$, $\bar{x}_1 = 1$ and $\lambda = 0.3$ for the first four vibration modes.

Similar results to those shown in Fig. 6 are depicted in Fig. 7 for the clamped-clamped beam, where the mass is attached at the mid-span, $\bar{x}_1 = 1/2$. As discussed earlier, the eccentricity of the mass positioned at the mid-span of a clamped-clamped beam does not affect the first and third frequencies, which is evident in Fig. 7(a) and (c). However, the eccentricity does influence the second and fourth frequencies, as demonstrated in Fig. 7(b) and (d). Furthermore, the effect of the aspect ratio on the second and fourth frequencies depends on the eccentricity of the attached mass. For instance, changing the aspect ratio from 0 to 0.3 alters the second frequency by 15% and 7% for the cases with eccentricities equal to 0 and 0.3, respectively.

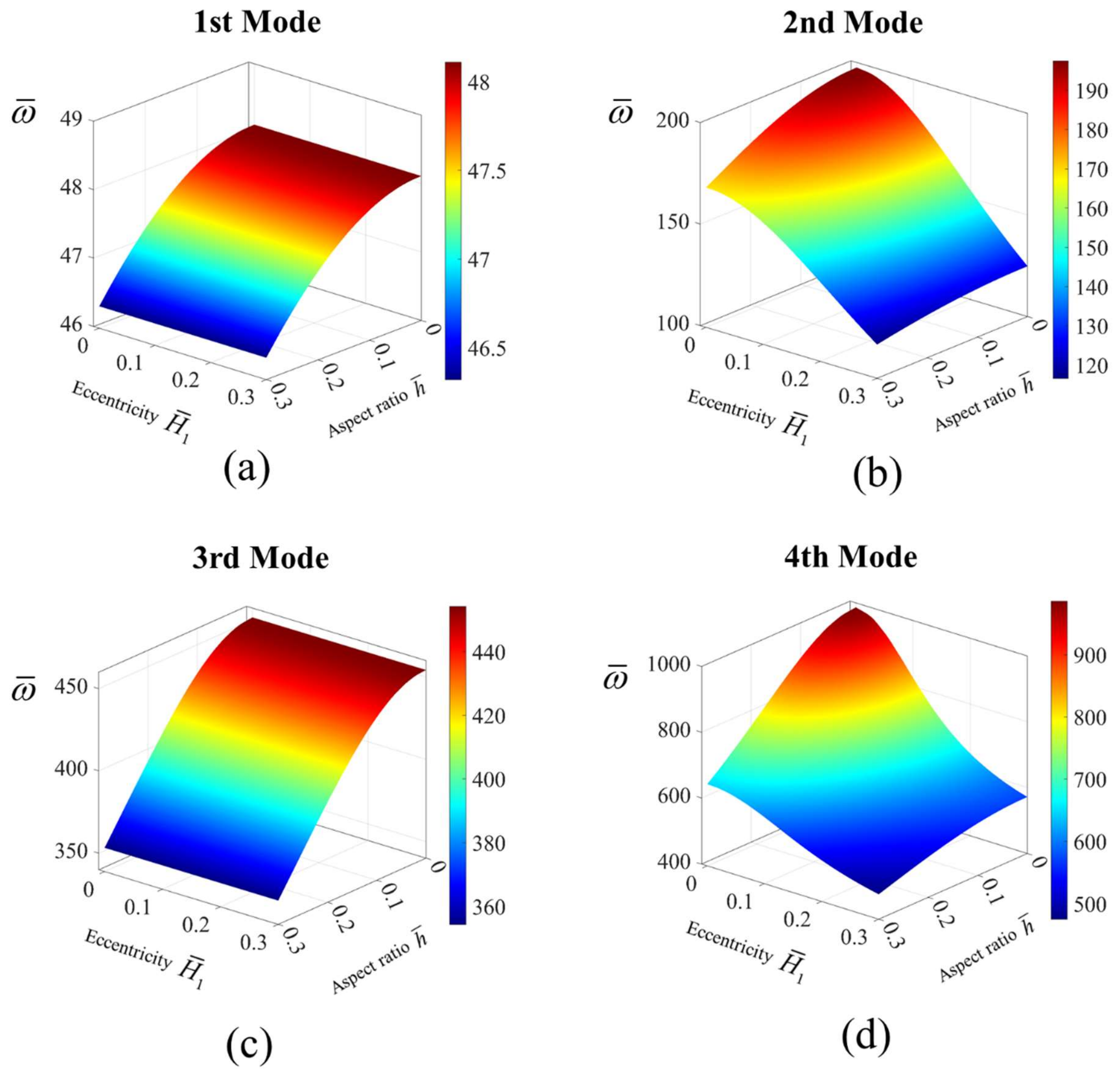

**Fig. 7** Frequencies vs. mass eccentricity and aspect ratio of the miniaturized clamped-clamped sensor with $\bar{M}_1 = 0.1$, $\bar{x}_1 = 1/2$ and $\lambda = 0.3$ for the first four vibration modes.

The eccentricity of an attached mass also affects the mode shapes of micro- and nanomechanical mass sensors. The first four vibration mode shapes of the nonlocal ($\lambda = 0.5$) cantilever sensor with an attached mass, $\bar{M}_1 = 0.1$, at the free end, $\bar{x}_1 = 1$, and $\bar{h} = 0.1$, are shown in Fig. 8. The mode shapes are presented for three different values of the eccentricity of the attached mass, $\bar{H}_1 = 0.1$, 0.2, and 0.3. The eccentricity of the attached mass does not change the first mode shape noticeably. This observation aligns with the findings in Fig. 4, where the first frequency of the micro- and nanocantilever mass sensors is less influenced by the eccentricity of the attached mass compared to the higher frequencies. However, the figure demonstrates a noticeable influence of the eccentricity of the attached mass on the higher mode shapes of the sensor. Specifically, as the eccentricity of the attached mass increases, the locations where the mode shapes exhibit maximum deflections are drawn slightly closer to the free end. Moreover, higher eccentricity also affects the positions of the nodes, where the deflection is zero, causing them to shift towards the free end as well. This effect is attributed to the increase in kinetic energy of the attached mass associated with axial velocities as the eccentricity increases. The heightened kinetic energy resulting from higher eccentricity alters the dynamic behavior of the sensor, including the shapes of the higher modes of vibration.

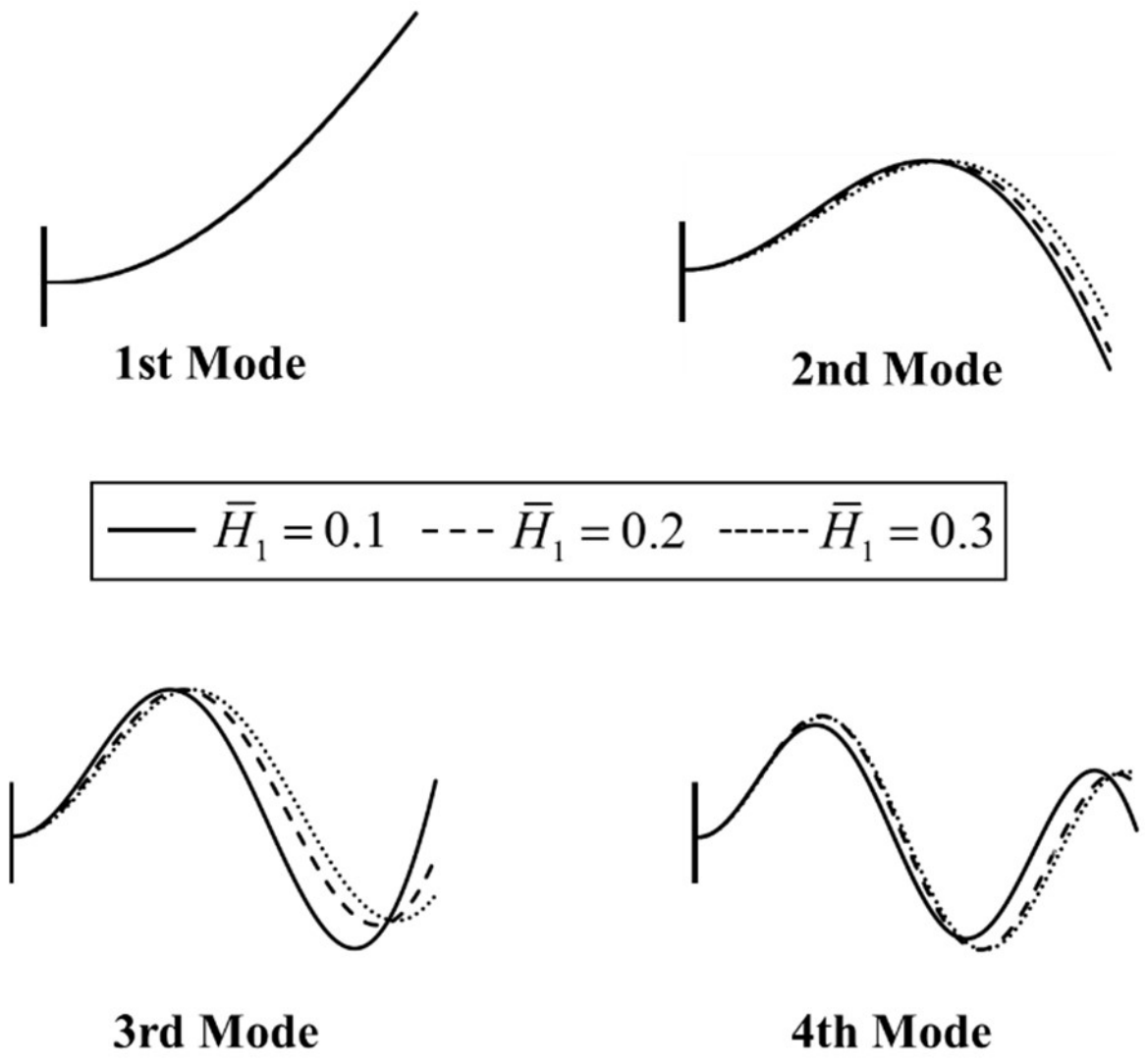

**Fig. 8** First four mode shapes of a nonlocal ( $\lambda = 0.5$ ) micro- or nanocantilever sensor with an attached mass, $\overline{M}_1 = 0.1$, at the free end, $\overline{x}_1 = 1$, and $\overline{h} = 0.1$, for $\overline{H}_1 = 0.1$, 0.2, and 0.3.

## 3.3. Size Effect

*3.3.1 Single mass*

The frequency percentage change in a cantilever micro- or nanosensor, due to the attachment of a mass at the free end, $\overline{x}_1 = 1$, is shown in Fig. 9, for the first four modes of vibration. The thickness-to-length ratio of the sensor is $\overline{h} = 0.1$, and the dimensionless distance between the mass center of the attached particle and the mid-thickness of the sensor is $\overline{H}_1 = 0.2$. Results are presented for the nonlocal parameter equal to $\lambda = 0$ (local sensor), 0.1, 0.2, 0.3, 0.4, and 0.5 on varying the dimensionless mass of the particle, $\overline{M}_1$.

In all cases, an increase in the dimensionless mass of the particle leads to a reduction in the natural frequencies. This reduction is particularly significant for sensors with higher values of the nonlocal parameter. In general, mass sensors constructed of materials with characteristic lengths comparable to their dimensions (nonlocal sensors) exhibit greater sensitivity to the presence of mass at the free end than sensors with negligible material characteristic lengths compared to the extrinsic dimensions (local sensors). This is because according to the stress-driven theory, as the dimensions of the sensor decrease, it becomes stiffer, leading to higher sensitivity to the presence of the tip mass. This enhanced sensitivity is generally more pronounced in the second mode of vibration and less noticeable in the fourth mode. For instance, the changes in the frequencies of the first, second, third, and fourth modes of vibration of the sensor with $\lambda = 0.5$, due to the presence of a particle with $\overline{M}_1 = 0.05$ at the free end, are, respectively, 17, 50, 31, and 8% higher than those of the sensor with $\lambda = 0$. Hence, overlooking size dependence can have a significant impact on the accuracy of the micro- or nanomechanical mass sensor and lead to misleading mass detections. Specifically, when the attached particle is positioned at the free end, disregarding size dependence leads to an overestimation of the particle's mass. This overestimation becomes more pronounced when the mass detection relies on the frequency shift of the second mode of vibration.

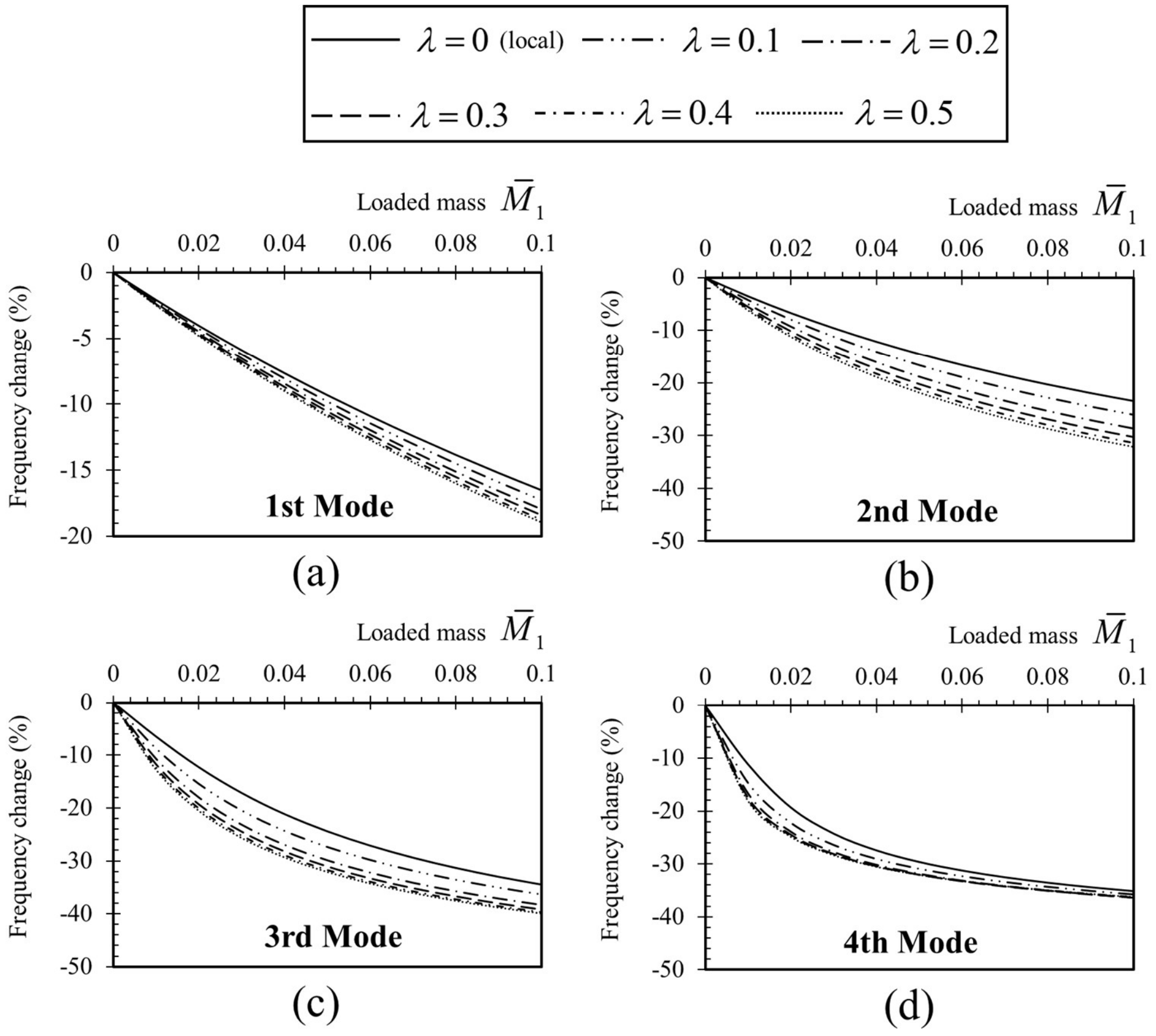

**Fig. 9** Frequency shifts related to the first four vibration modes in a miniaturized cantilever sensor with $\bar{h}=0.1$ and $\bar{H}_1=0.2$ after mass attachment at the free end.

The extent of overestimation or underestimation of the mass of the attached particle resulting from neglecting the size dependence varies depending on the location of the mass and the mode of vibration. This variability is illustrated in Fig. 10, which shows the percentage changes in frequency for the first four modes of vibration in a micro- or nanocantilever sensor due to the attachment of a mass at the mid-span. The illustrated data in the curves are associated with the sensor having $\bar{h}=0.1$, and $\bar{H}_1=0.2$. The results are presented for varying the nonlocal parameter, $\lambda$, and the dimensionless loaded mass, $\bar{M}_1$. When mass detection relies on the first three vibration modes, neglecting the size effect leads to an underestimation of the mass of the attached particle. This underestimation is more pronounced for the third vibration mode and diminishes for the first and second modes. However, when mass detection is based on the frequency shift of the fourth mode of vibration, neglecting the size effect results in overestimation.

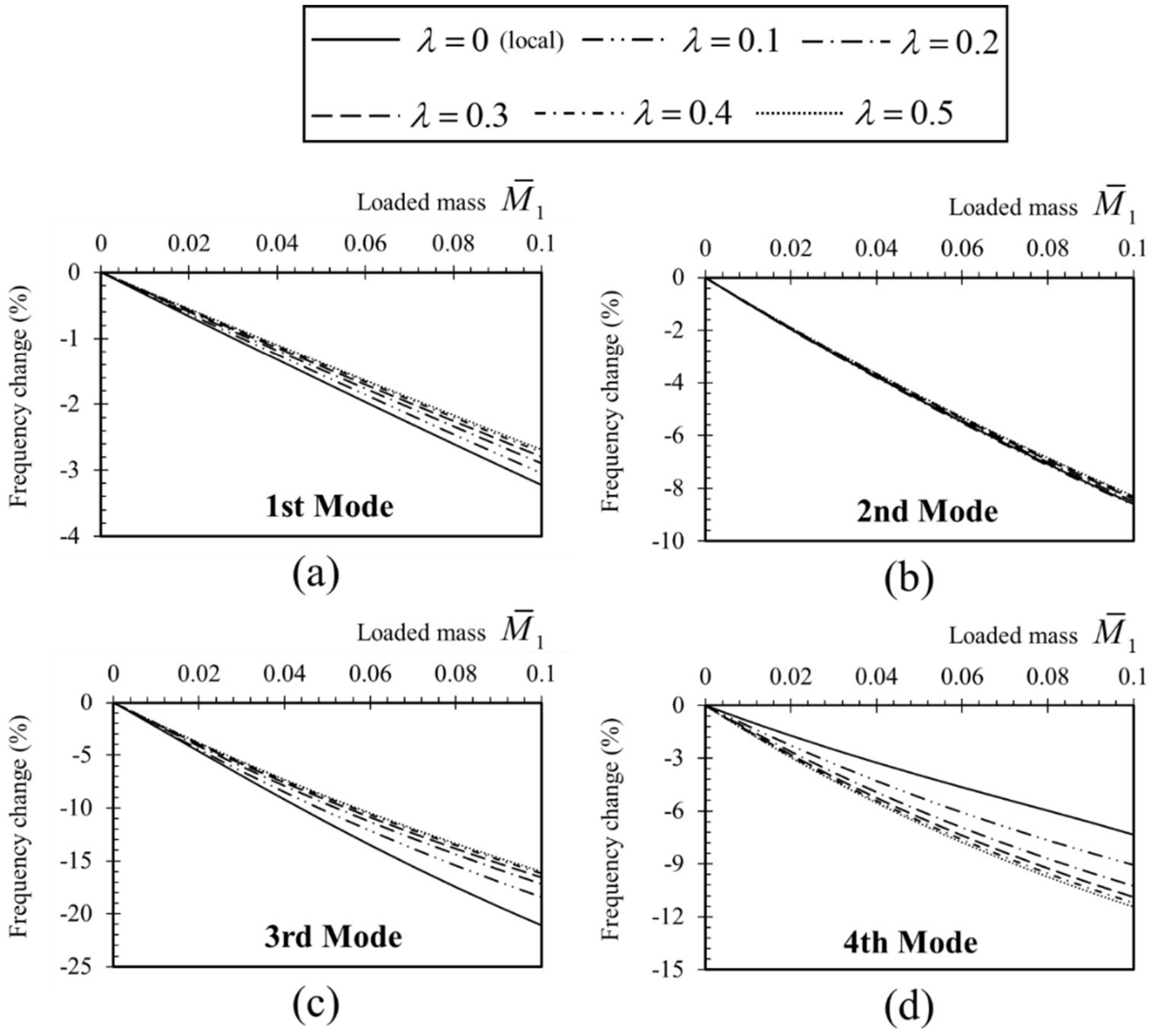

**Fig. 10** Frequency shifts related to the first four vibration modes in a miniaturized cantilever sensor with $\bar{h} = 0.1$ and $\bar{H}_1 = 0.2$ after mass attachment at the mid-span.

The frequency shifts in the first four vibration modes, induced by attaching a mass with $\bar{M}_1 = 0.1$ at different locations, are illustrated in Fig. 11 under the conditions of $\bar{h} = 0.1$, and $\bar{H}_1 = 0.2$, and while varying the nonlocal parameter, $\lambda$. In the first vibration mode, disregarding size dependence leads to underestimation and overestimation of the measured mass for attachments within intervals approximately equal to $\bar{x}_1 \leq 0.73$ and $\bar{x}_1 > 0.73$, respectively. Similar behavior is observed in the second mode, occurring at around $\bar{x}_1 = 0.72$. The third and fourth modes exhibit a more intricate pattern influenced by the attachment location. For instance, if the size effect is ignored, attaching a particle at about $\bar{x}_1 \leq 0.6$ results in a mass underestimation in the third mode, while attachment within $0.6 < \bar{x}_1 < 0.72$ leads to mass overestimation based on the frequency change in the third mode of vibration.

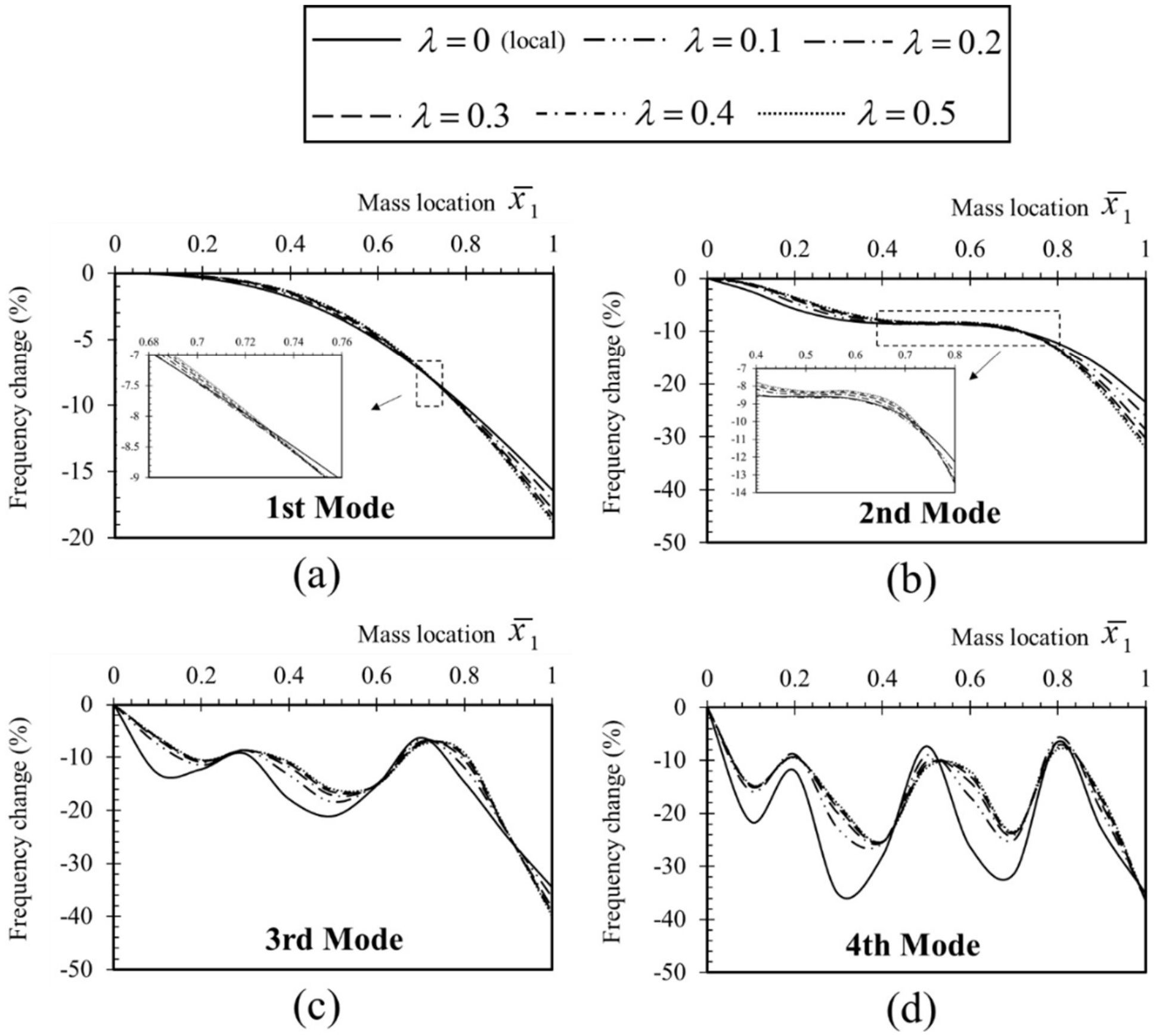

**Fig. 11** Frequency shifts related to the first four modes of vibration in a miniaturized cantilever sensor with $\bar{h} = 0.1$ and $\bar{H}_1 = 0.2$ after attachment of a mass with $\bar{M}_1 = 0.1$ at different locations.

The effect of neglecting the size dependence on the accuracy of the micro- or nanomechanical mass sensor depends also on the boundary conditions. The frequency percentage changes of the first four modes of vibration of a clamped-clamped sensor due to the attachment of a mass with $\bar{M}_1 = 0.1$ at different locations are shown in Fig. 12 for the case with $\bar{h} = 0.1$, and $\bar{H}_1 = 0.2$ on varying the nonlocal parameter, $\lambda$ . When the mass detection is based on the second, third, and fourth modes of vibration, neglecting the size effect generally results in underestimation. This is true also for the mass detection based on the first mode of vibration when the mass is located at distances within approximately $\bar{x}_1 \leq 0.25$ and $\bar{x}_1 \geq 0.75$ . However, if the mass is attached at any distance within the interval approximately equal to $0.25 < \bar{x}_1 < 0.75$ , neglecting size dependence results in the overestimation of the mass based on the frequency change of the first mode of vibration.

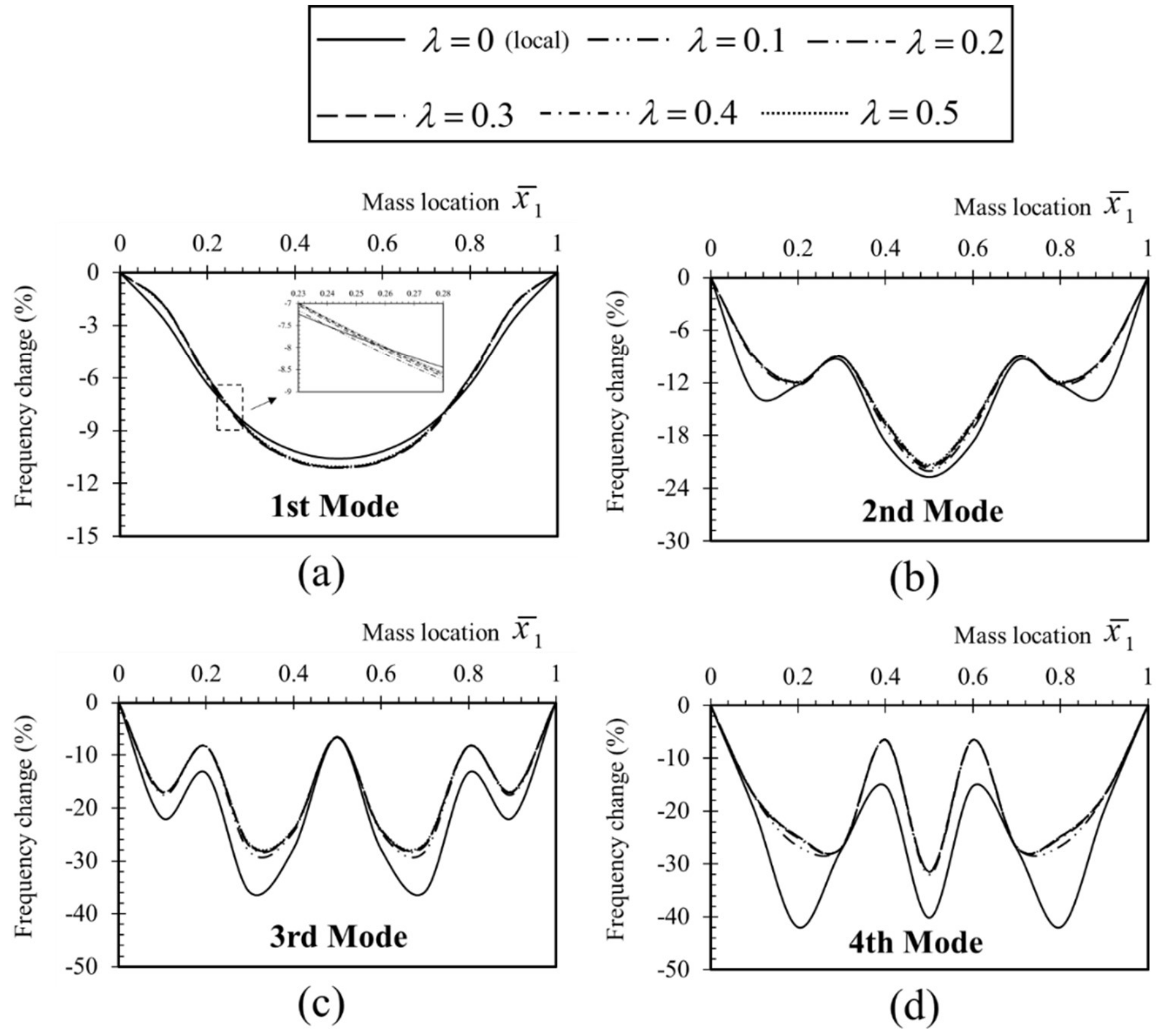

**Fig. 12** Frequency shifts related to the first four modes of vibration in a miniaturized clamped-clamped sensor with $\bar{h} = 0.1$ and $\bar{H}_1 = 0.2$ after attachment of a mass with $\bar{M}_1 = 0.1$ at different locations.

The size dependence also affects the mode shapes of the micro- and nanomechanical mass sensors. The first four vibration mode shapes of the local ($\lambda = 0$) and nonlocal ($\lambda = 0.5$) micro- or nanocantilever sensor with an attached mass, $\bar{M}_1 = 0.1$, at the free end, $\bar{x}_1 = 1$, $\bar{h} = 0.1$, and $\bar{H}_1 = 0.2$, are shown in Fig. 13. The mode shapes of a small-scale sensor (i.e. the nonlocal sensor) exhibit variations compared to those of a large-scale sensor. These changes include shifts in the location of maximum deflection. The influence of nonlocality on the fourth mode shape is comparatively subtle when compared to the other modes. This observation aligns with the findings in Fig. 9, where the frequency percentage changes in the fourth mode of vibration for both local and nonlocal sensors with $\bar{M}_1 = 0.1$ are closer to each other than those in the first three modes of vibration.

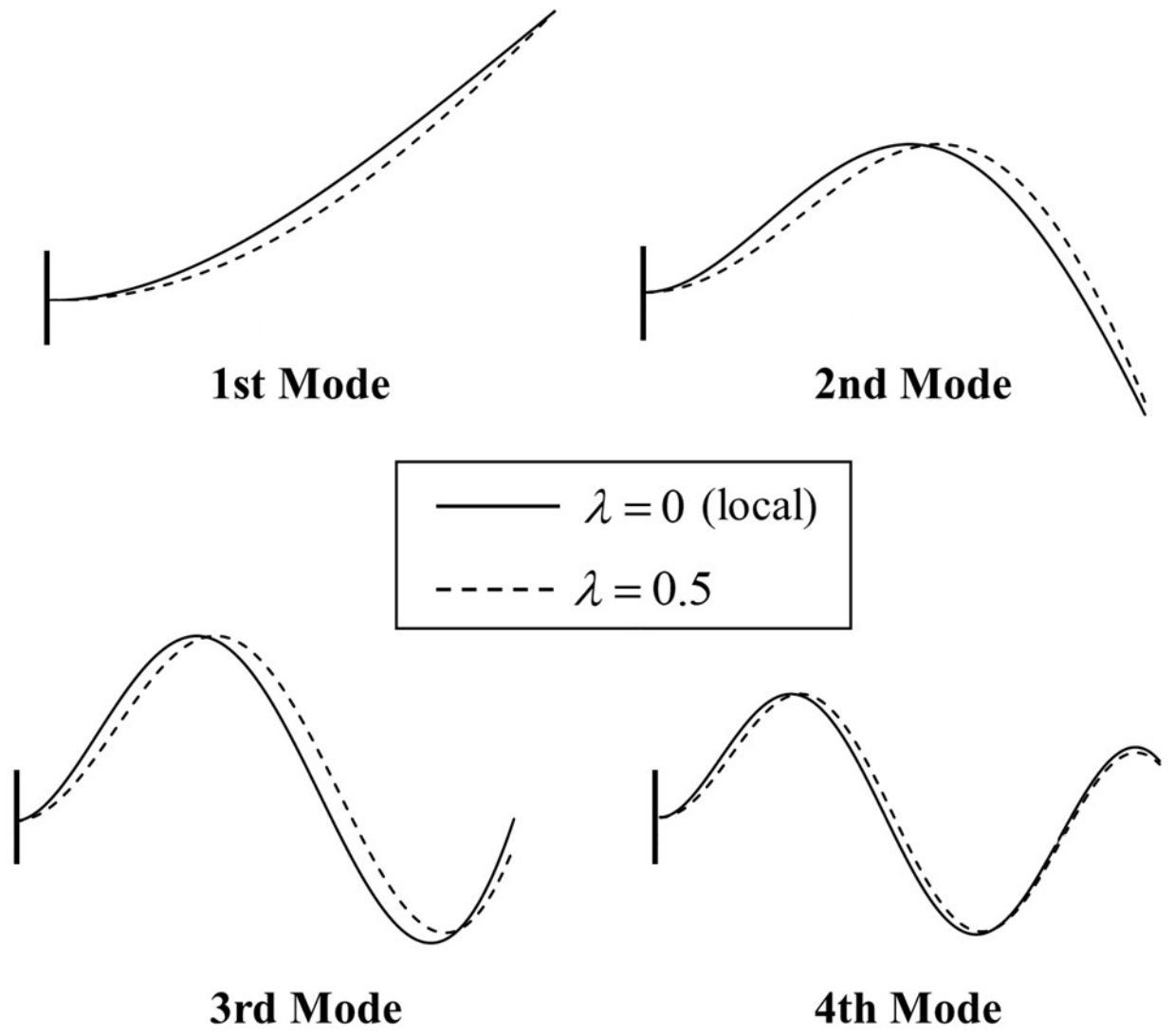

**Fig. 13** First four mode shapes of a local ( $\lambda = 0$ ) and nonlocal ( $\lambda = 0.5$ ) micro- or nanocantilever sensor with an attached mass, $\bar{M}_1 = 0.1$, at the free end, $\bar{x}_1 = 1$, $\bar{h} = 0.1$, and $\bar{H}_1 = 0.2$.

*3.3.2 Multiple masses*

The dimensionless natural frequencies of the first two modes of vibration of a clamped-clamped micro- or nanosensor with two attached masses $\bar{M}_1 = \bar{M}_2 = 0.15$, $\bar{h} = 0.1$, and $\bar{H}_1 = \bar{H}_2 = 0.2$ are shown in Fig. 14. Results are presented for $\bar{x}_1 = 0.25$ on varying the nonlocal parameter, $\lambda$ , and the location of the second mass, $\bar{x}_2$, from 0.3 to 0.95. For both modes of vibration, increasing the nonlocal parameter increases the sensitivity of the natural frequency to the location of the second mass. Increasing $\bar{x}_2$ results in a decrease in the natural frequency of the first mode of vibration for $\bar{x}_2 \leq 0.5$, and an increase when $\bar{x}_2 > 0.5$. As can be seen in the figure, the effect of the location of the second mass on the natural frequency of the second mode of vibration is more complex than that of the first mode of vibration.

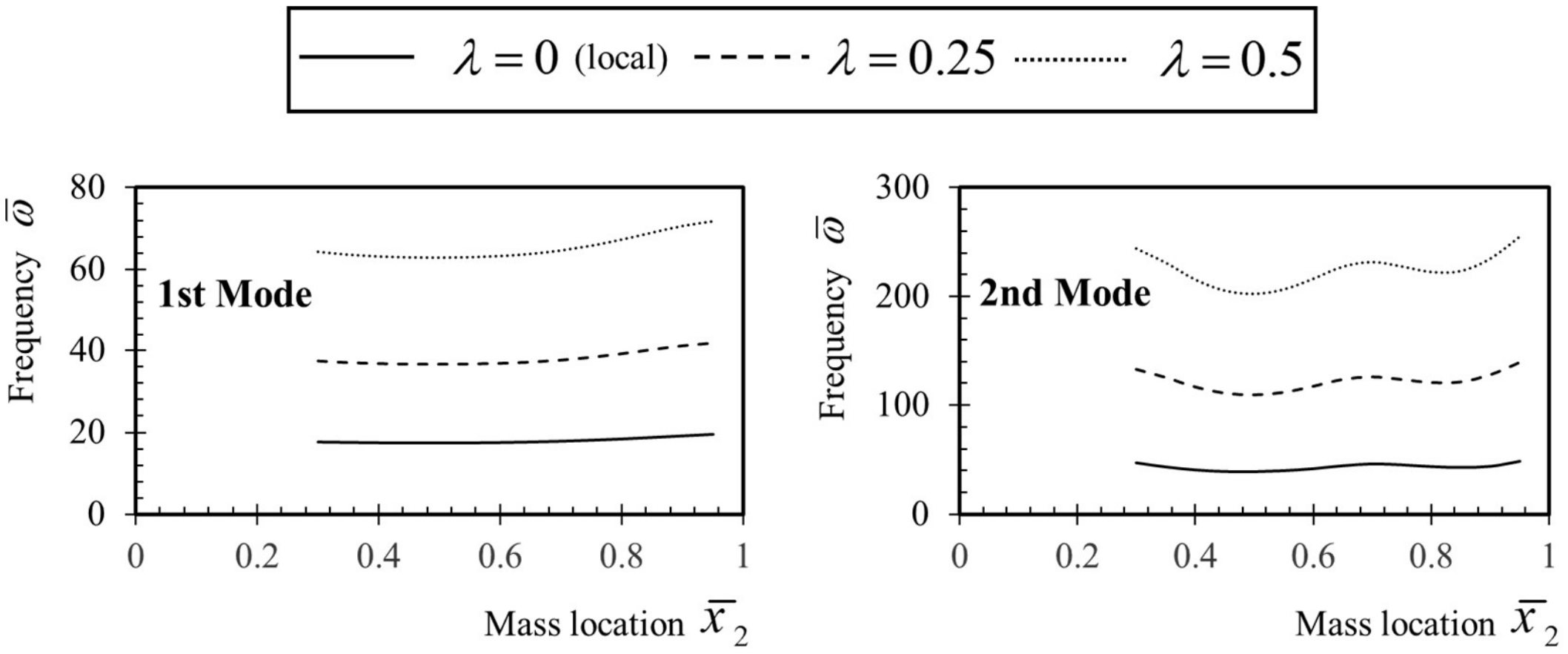

**Fig. 14** Dimensionless natural frequencies of the first two modes of vibration of a clamped-clamped micro- or nanosensor with two attached masses $\overline{M}_1 = \overline{M}_2 = 0.15$, $\overline{h} = 0.1$, $\overline{x}_1 = 0.25$ and $\overline{H}_1 = \overline{H}_2 = 0.2$.

The frequencies of a cantilever sensor in the presence of three attached masses are depicted in Fig. 15. The results refer to the case with the following parameters: $\overline{x}_1 = 0.3$, $\overline{x}_2 = 0.6$, $\overline{x}_3 = 1$, $\overline{h} = 0.1$, and $\overline{H}_1 = \overline{H}_2 = \overline{H}_3 = 0.2$. The results are shown for both local and nonlocal sensors, with variations in the dimensionless mass of each particle while maintaining the dimensionless masses of the other particles constant at 0.2. The figure highlights that the influence of the third mass at the free end dominates the impact of other masses on the frequencies. This is because the tip mass experiences the highest axial and transverse velocities, thus contributing significantly to the kinetic energy of the system. Therefore, the mass attached to the free end of the cantilever sensor plays a more critical role in determining the overall dynamic response and natural frequency of the system. This observation explains why the natural frequencies are nearly independent of the mass of the first particle, especially, and to a lesser extent, the second particle.

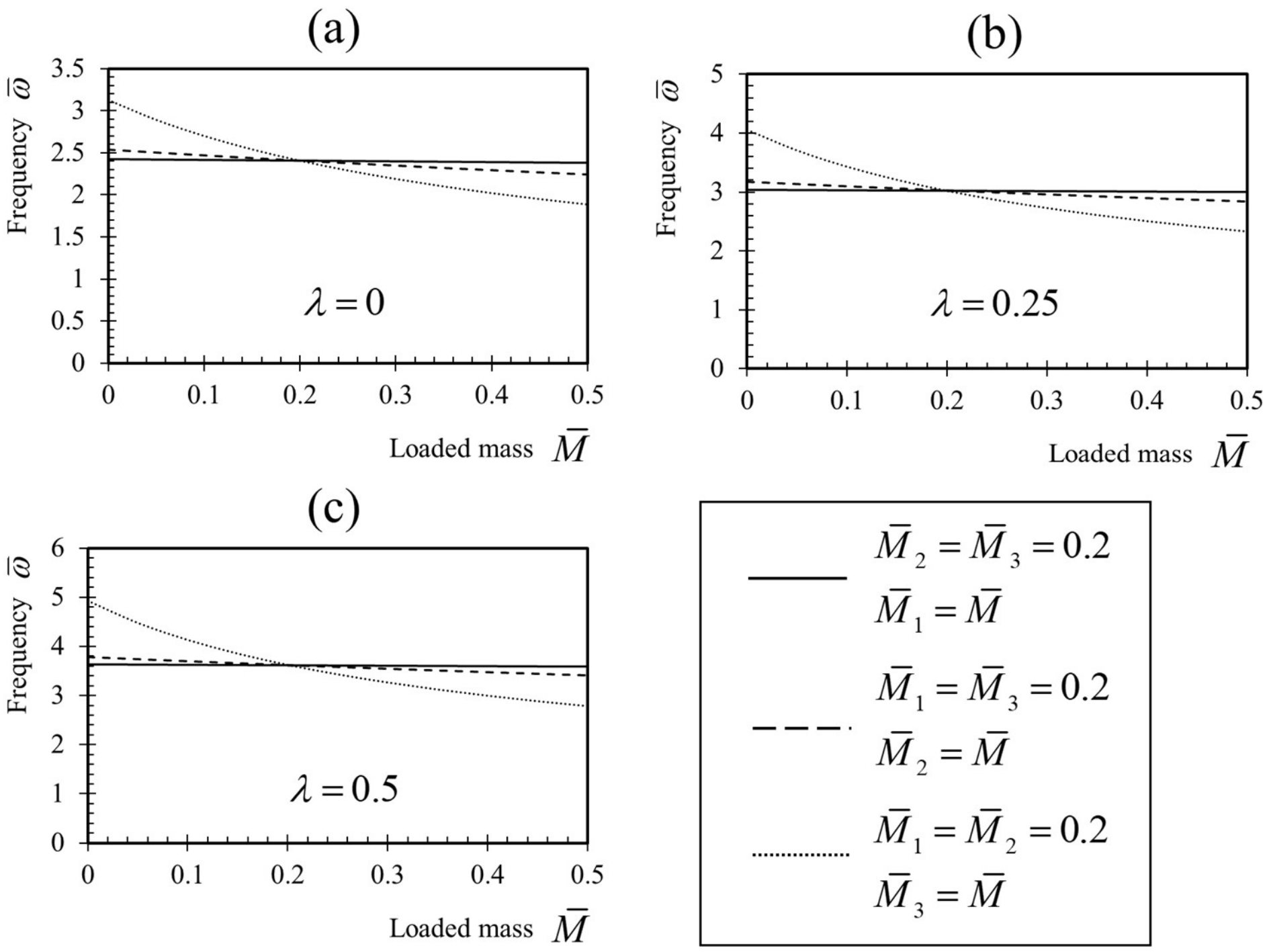

**Fig. 15** Natural frequencies of local and nonlocal cantilever micro- or nanosensors with three attached masses at $\bar{x}_1 = 0.3$, $\bar{x}_2 = 0.6$, and $\bar{x}_3 = 1$, $\bar{h} = 0.1$, and $\bar{H}_1 = \bar{H}_2 = \bar{H}_3 = 0.2$.

### 3.4. Inverse Problem

In many practical applications of ultrasensitive mass sensors, the entities may land at any position along the sensor. Therefore, in addition to the mass, the location of the attached particle may be also unknown. The problem of finding the location and the mass of an attached particle based on the natural frequency changes is considered here as the inverse problem. The changes in the first three frequencies of a micro- or nanocantilever mass sensor due to the attachment of a particle are shown in Fig. 16 for varying the dimensionless mass and location. The results are presented for the case of $\lambda = 0.2$, and $\bar{h} = \bar{H}_1 = 0.1$. The level sets showing the pairs $(\bar{x}_1, \bar{M}_1)$ for which the changes in the frequency are constant are presented in Fig. 16(d) for the first (thick lines) and second (thin lines) modes of vibration. It can be understood from Fig. 16(d) that the level sets corresponding to the first and second modes of vibration may intersect at more than one point. Therefore, the information regarding the frequency changes in the first and the second modes of vibration does not guarantee the uniqueness of the solution. In other words, there can be two possible solutions $(\bar{x}_1, \bar{M}_1)$ for which the desired changes in the first and the second frequencies are obtained. Two of these possible pairs of solutions are highlighted in Fig. 16(d). In this case, the information regarding the third frequency change would be necessary to identify the mass and location of the attached particle.

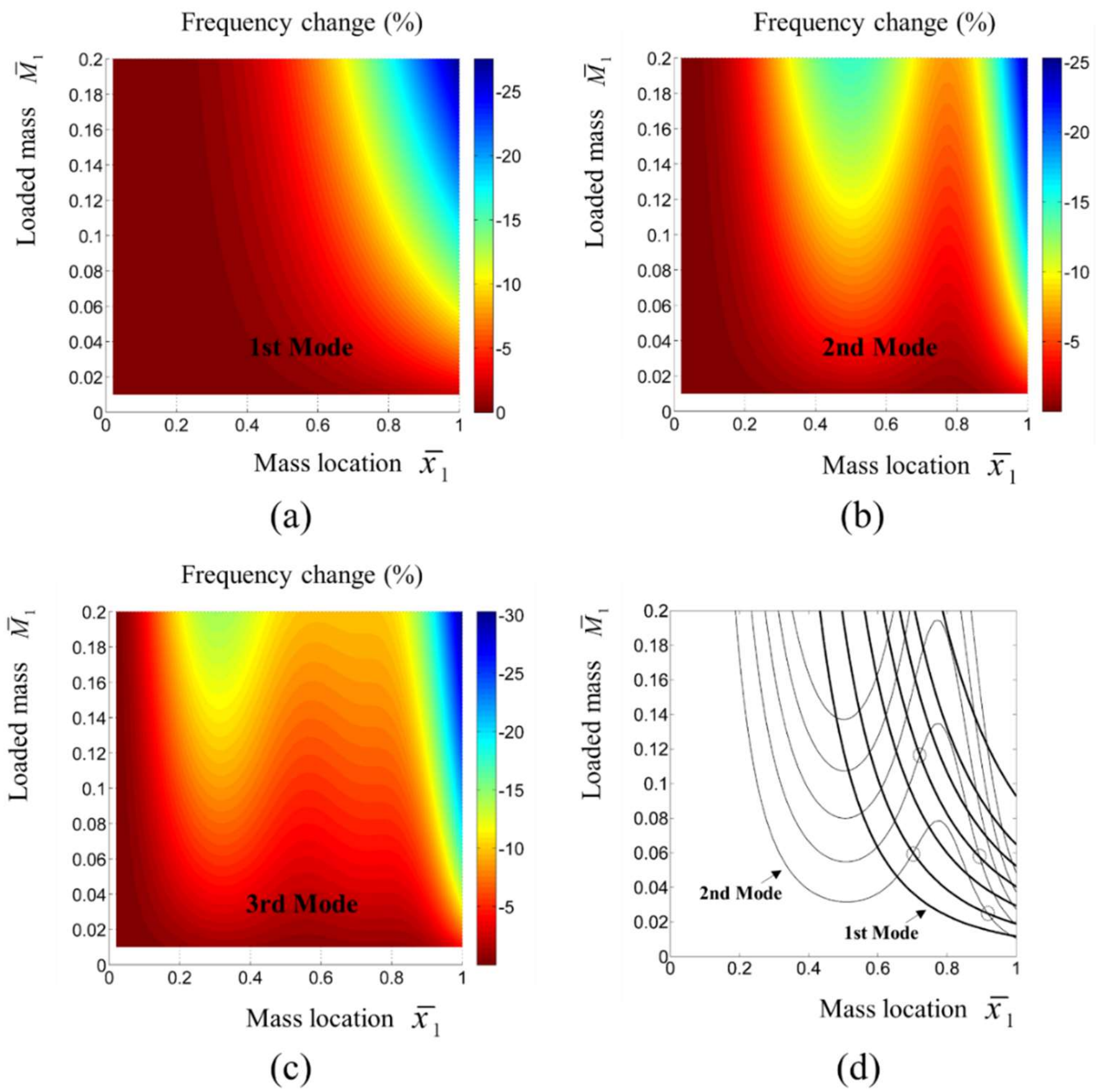

**Fig. 16** Frequency shifts of the (a) first, (b) second, and (c) third mode of vibration in a micro- or nanocantilever sensor due to the attachment of a mass on varying the dimensionless mass and location. The results are presented for the case with $\lambda = 0.2$, and $\bar{h} = \bar{H}_1 = 0.1$. (d) The level sets showing the pairs $\left(\bar{x}_1, \bar{M}_1\right)$ for which the changes in the frequency are constant. The level sets are shown for the first (thick lines) and second (thin lines) modes of vibration.

The size dependence may highly affect the solution of the inverse problem. This is shown in Fig. 17, where the level sets corresponding to -0.8 and -1.2% change in, respectively, the first (thick lines) and second (thin lines) modes of vibration are presented. The results correspond to the case with $\bar{h} = 0.1$, $\bar{H}_1 = 0.2$ and two different nonlocal parameters, $\lambda$, equal to 0 (local model, red lines) and 0.6 (black lines). The intersections of the level sets for the first and the second modes of vibration are marked. The locations of the intersections $(\bar{x}_1, \bar{M}_1)$ are equal to $(0.94, 0.0044)$ and $(0.84, 0.0055)$ for the local ( $\lambda = 0$ ) and nonlocal ( $\lambda = 0.6$ ) inverse problems, respectively. In other words, neglecting the size effect in sensors made of materials with characteristic lengths comparable to the sensor's dimensions would result in inaccurate detections in terms of both the mass and the location of the attached particle.

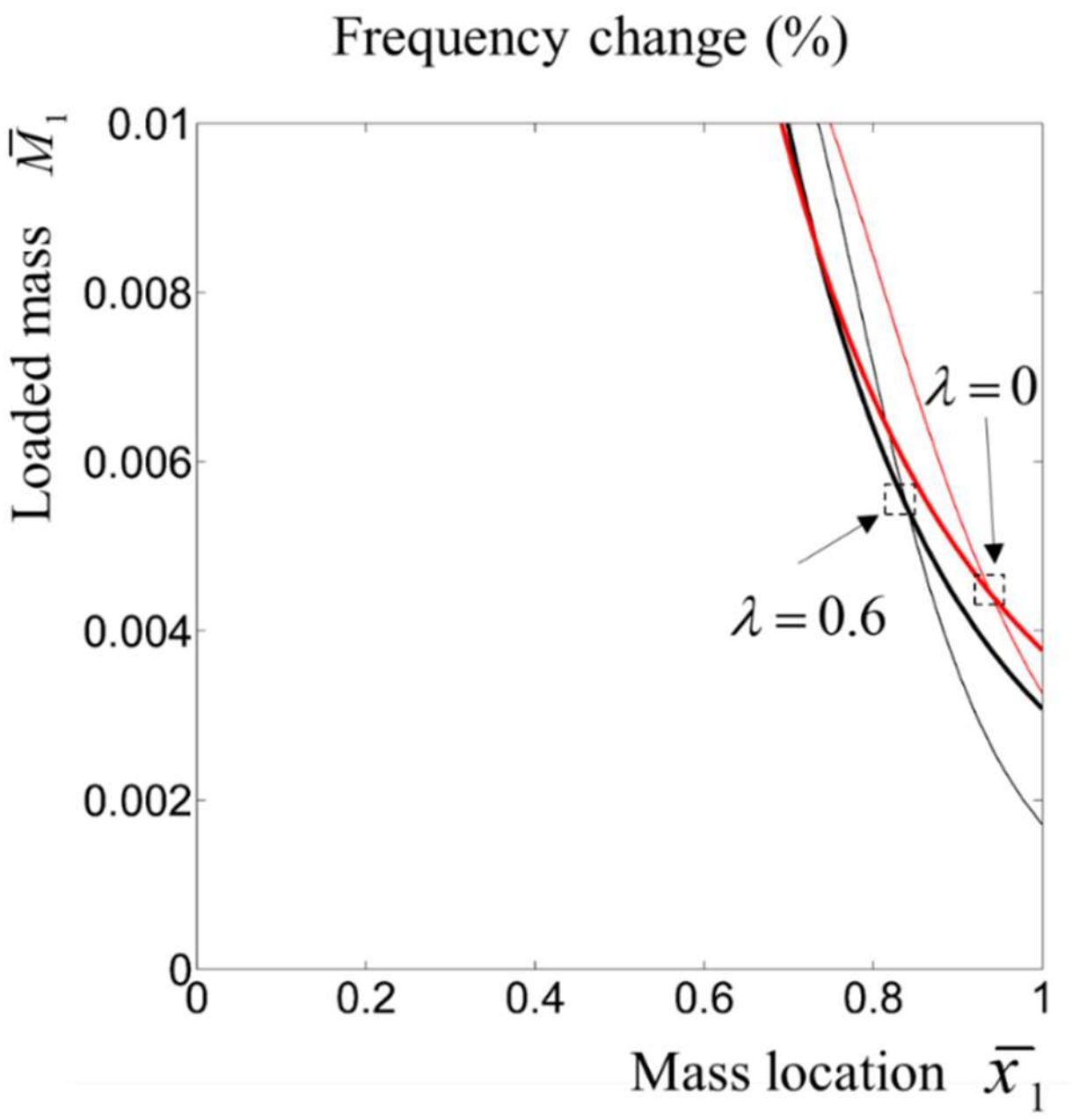

**Fig. 17** The level sets corresponding to -0.8 and -1.2% change in, respectively, the first (thick lines) and second (thin lines) modes of vibration in a micro- or nanocantilever sensor due to the attachment of a mass on varying the dimensionless mass and location. The results are presented for the case with $\bar{h} = 0.1$, $\bar{H}_1 = 0.2$ and two different nonlocal parameters, $\lambda$, equal to 0 (local model, red lines) and 0.6 (black lines). The intersections of the level sets for the first and the second modes of vibration are marked. The locations of the intersections $(\bar{x}_1, \bar{M}_1)$ are equal to $(0.94, 0.0044)$ and $(0.84, 0.0055)$ for, respectively, the local ( $\lambda = 0$ ) and nonlocal ( $\lambda = 0.6$ ) inverse problems.

## 4. Conclusions

The investigation of the size-dependent free transverse vibration of a micro- or nanobeam with any number of eccentric masses has been conducted employing the variational approach and the stress-driven nonlocal theory of elasticity. The kinematic field has been characterized using the Bernoulli-Euler beam theory. Dynamic equilibrium equations, along with variationally consistent boundary and continuity conditions at the cross-sections where the masses are attached, have been obtained through the Hamilton principle. The natural frequencies have been derived by solving the variationally consistent equations together with the higher-order constitutive boundary and continuity conditions. It has been demonstrated that the predictions of the model virtually coincide with the experimental and numerical results available in the literature. It has been shown that the eccentricity of the mass can greatly alter the frequencies depending on the location where the particle is attached. Furthermore, the size effect significantly influences the frequency shifts depending on the boundary conditions, the location of the particles, and the mode of vibration. The vibration mode shapes are also affected by the eccentricity of the attached masses and the small-scale size effects. In addition, the miniaturized beams are more sensitive to the presence of the attached particles compared to the large-scale beams. The inverse problem of finding an attached particle's location and mass based on the natural frequency changes has been considered. It has been shown that neglecting the size dependence may result in wrong detections of the mass and its location.

## Funding

Hossein Darban gratefully acknowledges the financial support provided by the National Science Centre (NCN) in Poland through the grant agreement No: UMO-2022/47/D/ST8/01348. 

## Declaration of Interests

The authors declare that they have no known competing financial interests or personal relationships that could have appeared to influence the work reported in this paper.

## Data Availability

Data will be made available on request.